\documentclass[]{pasj02}

\Received{}
\Accepted{}
 
\usepackage[switch,mathlines]{lineno}
\begin{document} 

\title{ 

Revisiting thermodynamics at the outskirts of the Perseus cluster with Suzaku: importance of modeling  the Hot Galactic gas}

\author{Kyoko \textsc{Matsushita}\altaffilmark{1}, Hayato \textsc{Sugiyama}\altaffilmark{1}, Masaki \textsc{Ueda}\altaffilmark{1}, Nobuhiro \textsc{Okabe}\altaffilmark{2,3,4}, Kotaro \textsc{Fukushima}\altaffilmark{1,5},  Shogo B. \textsc{Kobayashi}\altaffilmark{1}, Noriko Y. \textsc{Yamasaki}\altaffilmark{5}, and Kosuke \textsc{Sato}\altaffilmark{6}}
\altaffiltext{1}{Department of Physics, Tokyo University of Science, 1-3 Kagurazaka, Shinjuku-ku, Tokyo 162-8601, Japan}
\altaffiltext{2}{Physics Program, Graduate School of Advanced Science and Engineering, Hiroshima University, 1-3-1 Kagamiyama, Higashi-Hiroshima, Hiroshima 739-8526, Japan}
\altaffiltext{3}{Hiroshima Astrophysical Science Center, Hiroshima University, 1-3-1 Kagamiyama, Higashi-Hiroshima, Hiroshima 739-8526, Japan}
\altaffiltext{4}{Core Research for Energetic Universe, Hiroshima University, 1-3-1, Kagamiyama, Higashi-Hiroshima, Hiroshima 739-8526, Japan}
\altaffiltext{5}{ Institute of Space and Astronautical Science (ISAS), Japan Aerospace Exploration Agency (JAXA), 3-1-1 Yoshinodai, Chuo-ku,
Sagamihara, Kanagawa 252-5210, Japan}
\altaffiltext{6}{Department of Astrophysics and Atmospheric Sciences, Kyoto Sangyo University, Kyoto 603-8555, Japan}
\email{matusita@rs.tus.ac.jp}

\KeyWords{Clusters:X-rays--galaxies:clusters:intracluster medium--dark matter}

\maketitle

\begin{abstract}

The thermodynamic properties of the intracluster medium (ICM) at the outskirts of galaxy clusters provide valuable insights into the growth of the dark matter halo and the heating of the ICM.
Considering the results of the soft X-ray background study of non-cluster Suzaku fields,
 we revisit 65 Suzaku pointing observations of the Perseus cluster in eight azimuthal directions beyond  
 $\sim 1$ Mpc ($\sim$0.8 $r_{500}$).
A possible foreground component, whose spectrum is modeled as a 1 keV collisional ionization equilibrium plasma, 
significantly affects the temperature and density measurements of the ICM in cluster outskirts.
 
 The emission measures in the six arms are similar, showing that the radial slopes of temperature and density follow  $r^{-0.67\pm0.25}$ and $r^{-2.21\pm 0.06}$,   respectively. The radial pressure profile is close to the average profile measured by the Planck satellite. The resulting entropy slope is  $\propto r^{0.81\pm 0.25}$ , consistent with the theoretical slope of 1.1. The integrated gas fraction, the ratio of the integrated gas mass to the hydrostatic mass, is estimated to be   0.13$\pm$0.01 and 0.18$\pm$0.02   at $r_{500}$ and $r_{200}$, respectively, consistent with the cosmic baryon
fraction. These results suggest that the ICM at the cluster outskirts is quite regular and close to hydrostatic equilibrium.
 The remaining two arms show that the emission measure is higher by a factor of 1.5-2, possibly due to accretion from filaments from the large-scale structure. A sudden drop in the emission measure also occurs in a direction toward one of the filaments.

\end{abstract}

\section{Introduction}

Clusters of galaxies are the largest gravitationally bound objects in the Universe.
The cold dark matter cosmology predicts that 
clusters of galaxies continue to grow by accreting matter from their surroundings.
The accreting gas is expected to be heated by shock waves generated by the mass accretion and become an intracluster medium (ICM) \citep{Evrard1990}.
Such shocks increase the entropy of the ICM, and numerical simulations predict that the entropy increases with radius because the shock becomes stronger as a cluster grows \citep{Voit2005}.
When accretion occurs along a filament of the large-scale structure, the infalling gas
is expected to form a shock near the virial radius with a Mach number of 2--4, called the virial shock \citep{Evrard1990,simvirshock}.

Thanks to the low level of the non-X-ray background,
Suzaku enabled to study of the temperature and density of the ICM out to the virial radius of clusters of galaxies
(e.g.\cite{Bautz09},  \cite{Reiprich2009}, \cite{George2009}, \cite{Hoshino2010},    \cite{Kawaharada10}, \cite{Simionescu11}, \cite{Walker2012}, \cite{Sato2012}, \cite{Ichikawa13}, \cite{Okabe14}, \cite{Walker2019}).
The derived entropy profiles and baryon fractions, assuming hydrostatic equilibrium, differ significantly from the theoretical predictions: the observed entropy profiles tend to flatten beyond $\sim r_{500}$, and the baryon fraction sometimes exceeds the cosmic value.
\citet{Simionescu11}  proposed that inhomogeneous gas distributions (i.e., gas clumping) can lead to overestimates of the ICM density at large radii.
\citet{Kawaharada10} and \citet{Ichikawa13} compared the Suzaku results with weak-lensing mass estimates from Subaru observations and inferred a possible deviation from hydrostatic equilibrium.
They also found a possible correlation between the outskirts temperatures and the large-scale structure traced by galaxy distributions.

One limitation of Suzaku is its relatively poor spatial resolution, 2$^\prime$ (Half power diameter;  \cite{SuzakuXRT} ).
With the XMM satellite, possible clump candidates can be excluded thanks to its much better spatial resolution. 
However,  measuring the ICM temperature beyond $r_{500}$ is challenging due to the high non-X-ray background of XMM.
Measurements of the Sunyaev-Zel'dovich effect \citep{SZ}, where cosmic microwave background photons undergo inverse Compton scattering with the electrons in the ICM,  are complementary to X-ray observations.  
Combining the density profiles with {\it XMM} and the pressure profiles from   {\it Planck} data, \citet{XCOP2019} also studied the cluster outskirts and found that the temperature and density profiles of most clusters follow the predictions. 
They concluded that the ICM generally follows the predictions from simple gravitational collapse remarkably well.

The Perseus cluster is the brightest cluster in the sky.
A large-scale sloshing/swirling pattern in Perseus, characterized by X-ray bright low-entropy arcs and a spiral of cold front, has been mapped by mosaicking ROSAT PSPC, XMM-Newton, Suzaku, SRG/eROSITA data, revealing alternating features from the core out to hundreds of kiloparsecs and even $\sim$Mpc scales, consistent with minor-merger-induced gas sloshing \citep{Simionescu12, Walker2022, Churazov2025}.
\citet{Simionescu11} and \citet{Urban14} analyzed Suzaku data of this cluster 
and reported that the density profile of the ICM is flatter 
and the entropy profile is lower than expected from numerical simulations.
Near the virial radius ($\sim r_{200}$) towards the northwest, 
projected temperature and emission measures show discontinuities 
\citep{Urban14, Zhu2021}.
With XMM-Newton observations, \citet{Walker2022} found two 
X-ray surface brightness edges at 1.2 and 1.7 Mpc to the west and interpreted them as cold fronts caused by some accretion of smaller objects.

Since the X-ray emission from the cluster outskirts is very faint, the modeling of the background is crucial for the study of the ICM.  
The soft X-ray background below $\sim $ 1 keV is usually modeled by a sum of the Local Hot Bubble (LHB) and the Milky Way Halo (MWH).
The spectra of the LHB and MWH are modeled with collisional ionization equilibrium  (CIE) plasma at a temperature of 0.1 keV and 0.2--0.3 keV, respectively (e.g., \cite{Henley13}, \cite{Nakashima18}).
\citet{Yoshino09} found excess emissions at 0.7--1 keV in spectra of some regions without clusters and bright X-ray sources.
\citet{Sekiya14a}, \citet{Nakashima18}, \citet{Gupta2021}, and \citet{Gupta2022} also reported the presence of the 0.6-1.0 keV component in some regions observed with Suzaku. \citet{Halo22} also detected a similar temperature component with the HaloSat survey over 85\% of the sky at $|b|>30^\circ$. This component is particularly bright in the direction of the eROSITA bubble \citep{erositabubble}, which extends from the Galactic Center to the north and south.
With some Suzaku observations of the cluster outskirts beyond the virial radius, a similar spectral component that is modeled with a 0.6--0.9 keV CIE plasma was also detected (e.g., \cite{Ichikawa13}, \cite{Urban14}).
This spectral component can bias temperature and density measurements of the ICM in the cluster outskirts.

\citet{Paper I} and \citet{SugiyamaMWH2} analyzed Suzaku data of 130 observations without bright sources and clusters of galaxies at $75^\circ < l< 285^\circ, |b| > 15^\circ$. 
\citet{Paper I} reported that a spectral model including a hot CIE component with a temperature of 0.8--1.2 keV, in addition to the standard background model, reproduces the spectra of a significant fraction of the observations.
They also reported a strong correlation between the MWH emission measure and the sunspot number, suggesting that emissions associated with the solar activity, O\,\emissiontype {VII} He$\alpha$ possibly from the heliospheric SWCX, significantly contaminate the MWH component.  
For the observations around the solar minimum, at $105^\circ < l< 255^\circ$ and $|b|>35^\circ$,
the temperatures ($\sim 0.2$ keV) and emission measures of the MWH component are pretty uniform.
The plasma at the virial temperature of the Milky Way possibly fills the halo in near hydrostatic equilibrium.
\citet{SugiyamaMWH2} presented the result of the 0.8--1.2 keV emission component (the Hot Galactic, hereafter HG, component), including the same analysis as in \citet{Paper I}.
The emission measures of this component vary by at least one order of magnitude and tend to be higher at the lower galactic latitudes. 
The regions with the brightest HG component are concentrated towards the Orion-Eridanus Superbubble \citep{Reynolds79}, possibly blown from the Ori OB1 association.
They discussed possible relations between the hot bubbles and stellar feedback.

In this paper, we analyze Suzaku data beyond   $\sim $ 1 Mpc (0.8 $r_{500}$; section \ref{sec:hm})   of the Perseus cluster, taking into account the results of the soft X-ray background studies of \citet{Paper I} and 
 \citet{SugiyamaMWH2}.    
This paper is organized as follows.
In section 2, we describe the observations and data reduction.
In section 3, we present the results of the spectral fitting.
In section 4, we discuss the results.
In this paper,   we use the solar abundance table by Lodders et al. (2003).
We assume a $\Lambda$ CDM cosmology with $\Omega_{\rm m}=0.3$, $\Omega_\Lambda=0.7$, and  $H_0 = 70 ~\rm{km~ s^{-1} ~Mpc^{-1}}$.
We adopt the redshift of the Perseus cluster to be 0.018 \citep{redshift}.
Errors are reported at the 68\% confidence level unless otherwise noted.

\section{Observations and data reduction}

\begin{figure*}
          \centerline{\includegraphics[width=18cm]{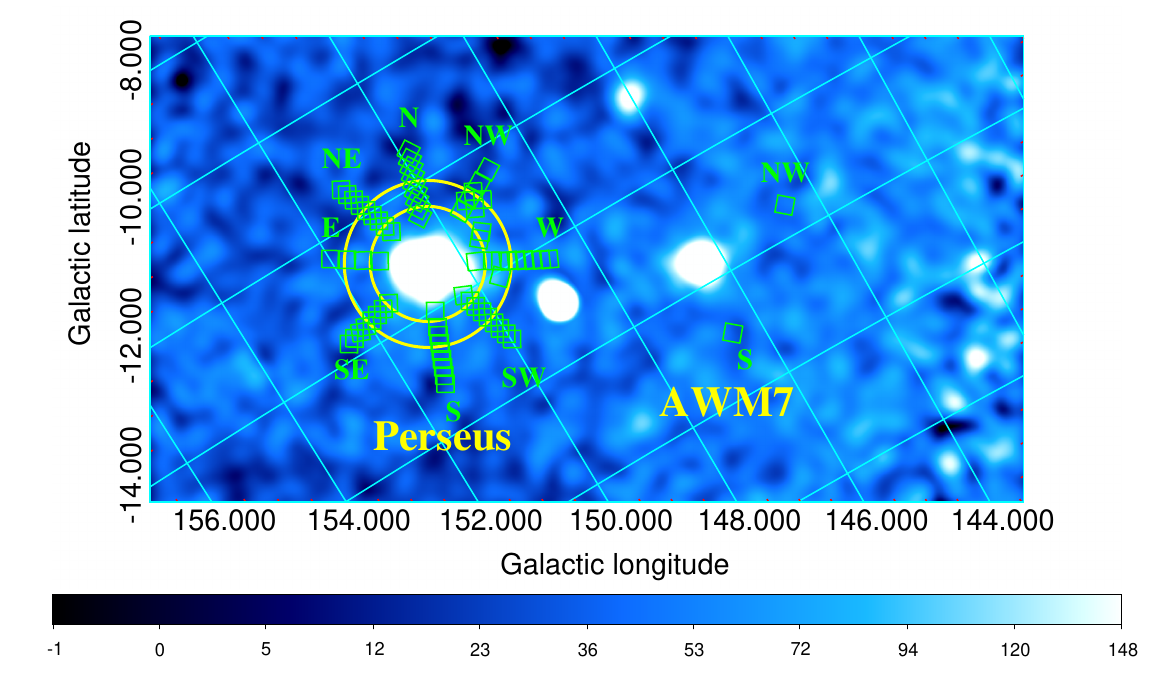}}
         \caption{The ROSAT soft-X-ray all-sky survey map in the 0.44--1.01 keV energy band \citep{Snowden1997} around the Perseus and AWM7 clusters. The squares show the XIS FOVs analyzed in this study.   The inner and outer circles correspond to $r_{500}$ and $r_{200}$, respectively.  
         {Alt text: Soft X-ray map in Galactic coordinates showing the Perseus and AWM7 clusters.  }}\label{fig:clusterimages}
                  \end{figure*}

We analyzed XIS data of the Perseus cluster.
The XIS detectors have a field of view (FOV) of $\sim 18^\prime\times 18^\prime$.
We selected 65 pointing observations beyond  $r\sim$ 1 Mpc ($\sim$0.8 $r_{500}$).  
Here, $r$ is the angular distance from the cluster center to the pointing position of each observation and 
 $r_{500}$  is the radius of a sphere whose mean total enclosed density is 500 times the critical density of the Universe. For the Perseus cluster, we estimated $r_{500}=59^\prime$ or 1.30 Mpc (section \ref{sec:hm}).
We also analyzed two outermost regions (three observations) of the AWM7 cluster observed with Suzaku.
The observation log is summarized in appendix \ref{appendix:observation_info}.
The pointing positions are plotted in figure \ref{fig:clusterimages} on the ROSAT all-sky survey map by \citet{Snowden1997}. We analyzed data belonging to eight arms towards N, NE, E, SE, S, SW, W, and NW.
The E and NW arms were observed in 2009, and the others in 2010 and 2011.
The four pointings in the NW arm were obtained in 2014 with a different roll angle from the overlap pointings observed in 2009. 
We added three observations at $r=$60$^\prime$-63$^\prime$ and 76$^\prime$ near the W arm.
The Perseus and AWM7 clusters belong to the Perseus-Pisces supercluster \citep{PPSC}.
The AWM7 cluster is located further out of the W arm, on a filament of the large-scale structure extending from the Perseus cluster.  Several clusters are aligned to the west and northwest of the Perseus cluster.  The K-band surface brightness map of galaxies also shows two clear filamentary structures in these directions   \citep{galaxies, Churazov2025}.  
The six arms, excluding the NW and W arms, are hereafter referred to as the relaxed arms.

The XIS has four CCD sensors. XIS0, 2, and 3 contain front-illuminated (FI) CCDs and
XIS1 has a back-illuminated (BI) CCD.
Since the XIS2 detector was lost in November 2006,  we analyzed the data from the three remaining CCDs.
Data obtained with 3$\times$3 and 5$\times$5 editing modes were merged.
We used the same filtering and point source removal method as in \citet{Paper I}.
We excluded the time ranges when the count rate exceeds three $\sigma$, in the same way as in \citet{Paper I}.

\section{Analysis and results}
\label{sec:ana}

We extracted a spectrum over the FOV of each XIS detector from each observation.
We rebinned each spectrum to at least one count per bin and used the extended C-statistic \citep{Cash79}.
We used {\it xissimarfgen}   \citep{Ishisaki2007}   to create auxiliary response files, assuming uniform emission from a circular region with a radius of 20$^\prime$.
We used XSPEC \citep{xspec} version 12.13.1. 
We fitted the XIS0, 1, and 3 spectra of the individual observations simultaneously with sky X-ray and non-X-ray background (NXB) components,
with the NXB
treated as an additive model in the spectral fitting rather than being subtracted beforehand.
The details of the NXB modeling are shown in Appendix \ref{sec:nxb}.
Here, we used the energy ranges of  0.4--11.5 keV for the BI spectra and 0.5--13.0 keV for the FI spectra.
 We used APEC (Astrophysical Plasma Emission Code, \cite{Smith01}, \cite{Foster12}) with AtomDB version 3.0.9 to model a CIE plasma.

\subsection{Spectral fitting of the background regions}

\begin{figure*}
  \centerline{
\includegraphics[width=4.5cm]{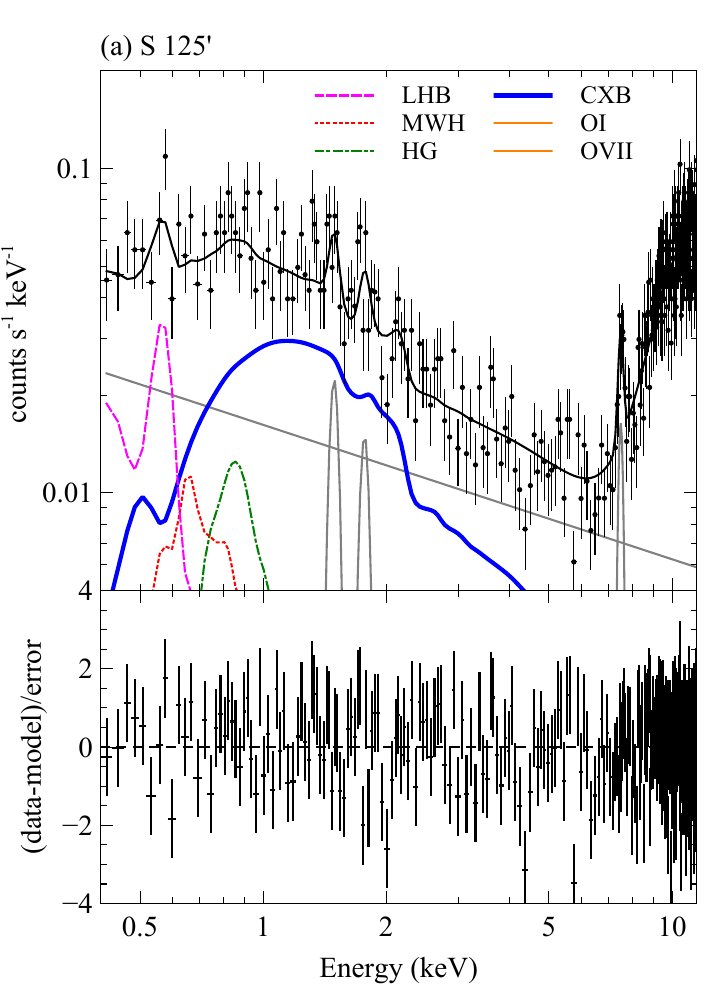}
\includegraphics[width=4.5cm]{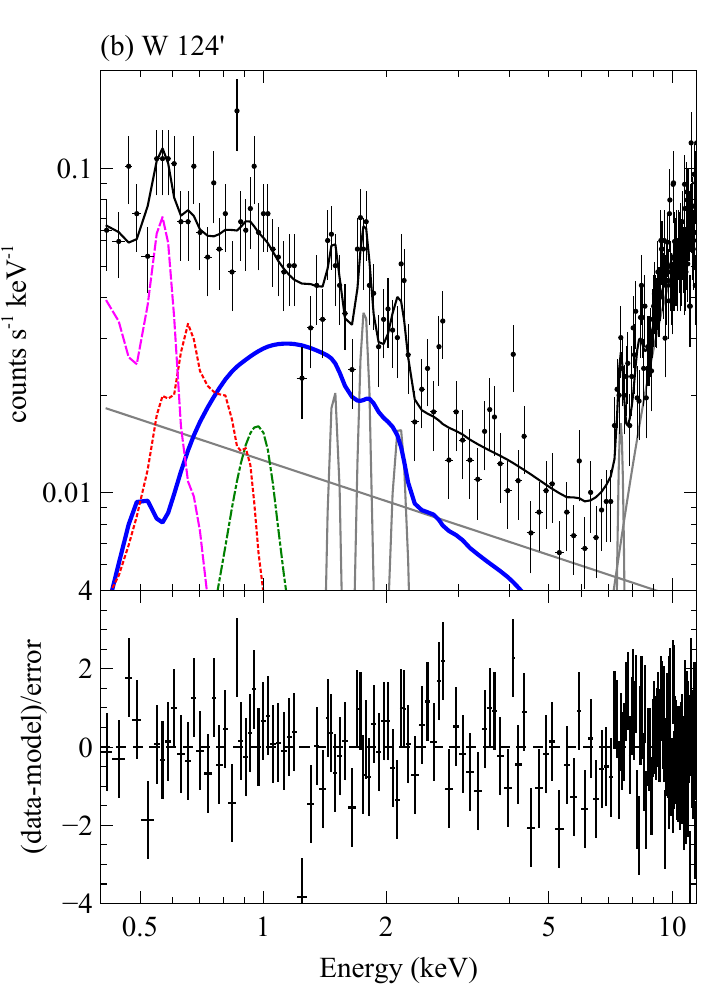}
\includegraphics[width=4.5cm]{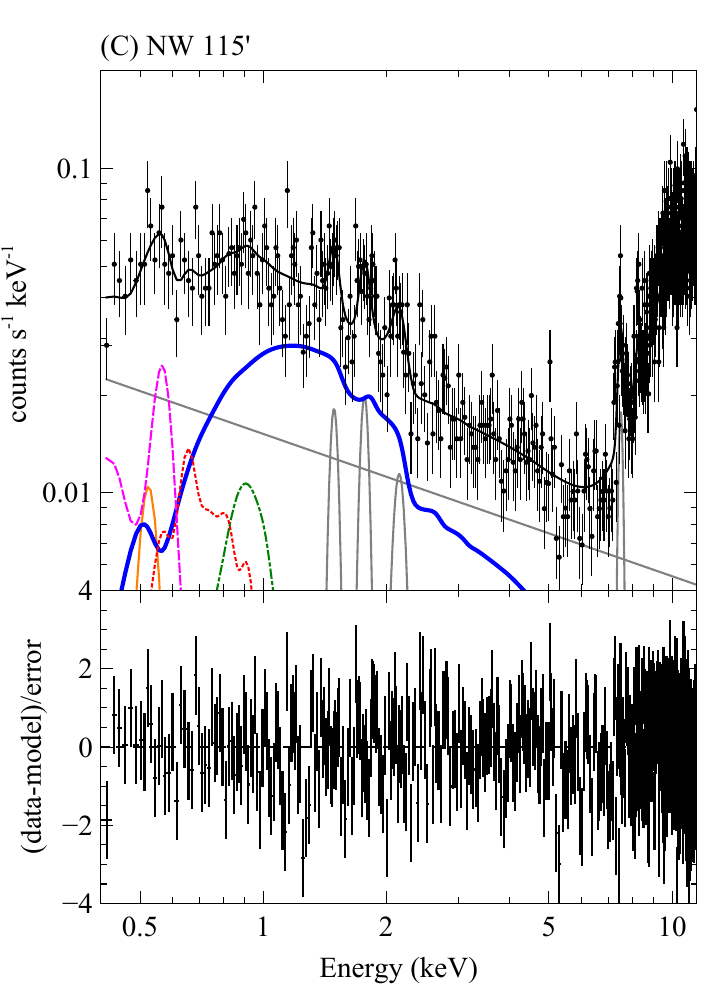}
\includegraphics[width=4.5cm]{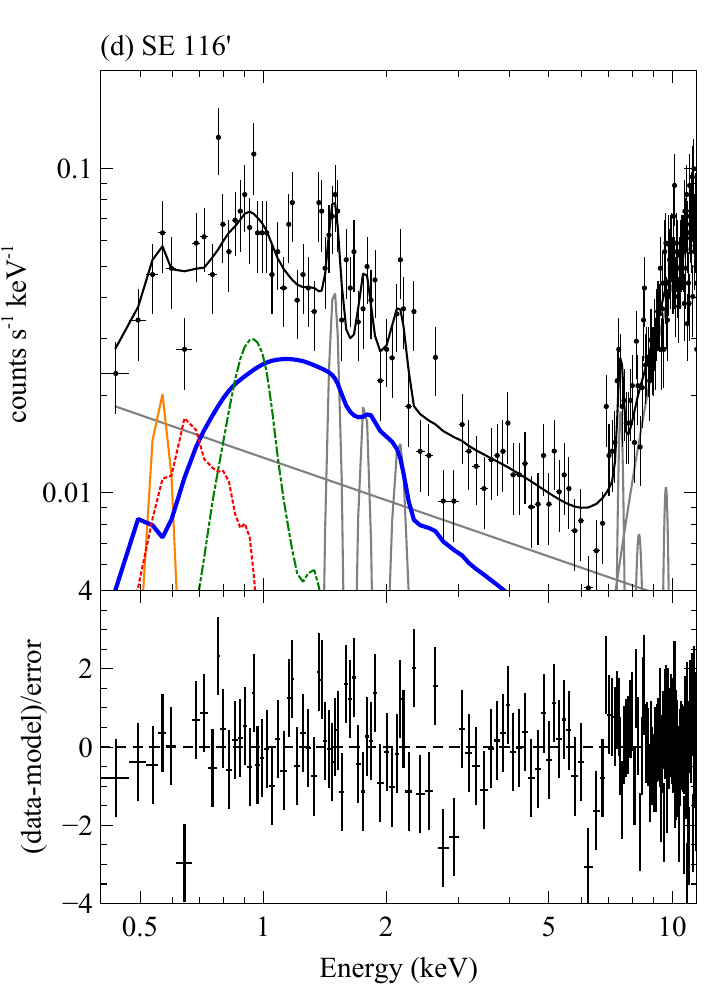}
}

\caption{The representative XIS1 spectra (upper panels) of the Perseus cluster at  $\sim 125^\prime$ in the S (a), and W (b) arms and at $\sim 116^\prime$ for the NW (c) and SE (d) arms, fitted with the background model and the residuals (lower panels). The contributions from the MWH (dotted red), the LHB (dashed magenta), the HG (dot-dashed green), the CXB (thick solid blue), O\,\emissiontype {I} (solid orange), and O\,\emissiontype {VII} (solid orange) are also shown. The contributions of the NXB components are shown as the gray solid lines.
 
{Alt text: Four subfigures, each consisting of two panels: the upper panels show the representative XIS1 spectra of the Perseus cluster, and the lower panels display the residuals.  Each subfigure corresponds to a different arm direction (S, W, NW, SE) at radii of approximately 116$^\prime$--125$^\prime$.}  
}
\label{fig:bgd}
\end{figure*}

\begin{figure}
  \centerline{\includegraphics[width=8cm]{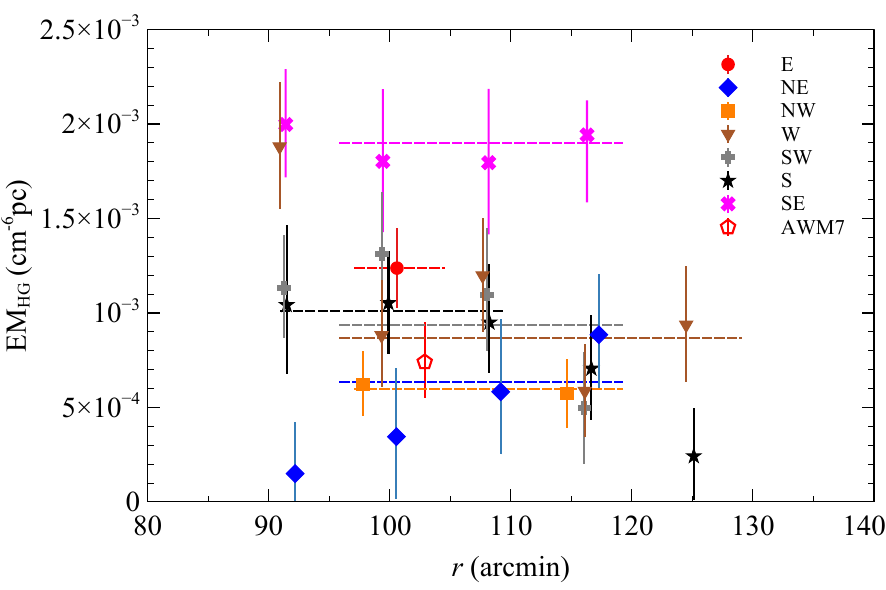}}

\caption{The radial profiles of EM$_{\rm HG}$ beyond   90$^\prime$   for the E (the red filled circle), NE (the blue filled diamonds), 
NW (the orange-filled squares), W (the brown downward triangles), SW (the gray pluses), S (the black stars), SE (the magenta crosses) arms,   and the NW background region of the AWM7 cluster (the red open pentagon).  Each dashed line represents the average value over the radial range covered by the corresponding data points with the same color.
 
{Alt text: Plot showing the radial profiles of EM$_{\rm HG}$ beyond 90$^\prime$ for multiple arms of the Perseus cluster and the NW background region of AWM7. A distinct symbol and color represent each arm. } }
\label{fig:08}
\end{figure}

Our background model consists of a power-law component for the cosmic X-ray background (CXB) and three CIE components ({\it apec} model in XSPEC) representing the LHB, MWH, and HG. Except for the LHB, these components are subject to the photoelectric absorption fixed at the Galactic value at each pointing position \citep{HI16}.
We added two Gaussians with fixed central energies at 0.525 keV and 0.561 keV to model the O\,\emissiontype {I}  K$\alpha$ line \citep{Sekiya14OI} and the O\,\emissiontype {VII}  He$\alpha$ line \citep{Paper I}, respectively, which sometimes contaminate the spectra near the solar maximum.
The photon index of the power-law component was fixed at 1.4.
The normalization of the CXB component was allowed to vary within a limited range of    9.0-10.4  
photons $\rm{cm^{-2}s^{-1}keV^{-1}sr^{-1}}$ at 1 keV(appendix \ref{back} for detail).
Outside the eROSITA bubble and using the data around the solar minimum, \citet{Paper I} found that the MWH component's temperature is uniform and consistent with their median value at 0.22 keV.
We fixed the temperatures of the LHB and MWH at 0.1 keV and 0.22 keV, respectively, and set their abundances to 1 solar. 
Further details of the soft X-ray background modeling are given in appendix \ref{soft}.

The temperatures of the HG component (hereafter $kT_{\rm HG}$) reported by \citet{SugiyamaMWH2} are typically $\sim$0.8~keV, although some fields exhibit higher values of 1$\sim$1.2~keV. 
\citet{Urban14} included a CIE component with a temperature of 0.6~keV in their analysis of the \textit{Suzaku} spectra of the Perseus cluster.
Because our soft X-ray background model and the fitted energy ranges differ from those of \citet{Urban14}, we first estimated $kT_{\rm HG}$ around Perseus by fitting spectra beyond $115^\prime$ from the Perseus center and two fields (three observations) in the outskirts of the AWM~7 cluster. 
  Here, the metal abundance of the HG component was fixed at 1 solar.  .
For AWM~7, the S and NW pointings yield  $kT_{\rm HG}=1.08\pm0.08$~keV (average of the two S observations) and $0.99\pm0.15$~keV, respectively.  
The outskirts of Perseus are consistent with $kT_{\rm HG}\simeq1$~keV. 
Based on these results, we adopt $kT_{\rm HG}=1.0$~keV and keep it fixed in the subsequent analysis.
We note that, without including an additional O\,VII He$\alpha$ line, and when the background temperatures are allowed to vary, the best-fit temperatures of the Milky Way halo (MWH) and HG components in the NW background region of AWM~7 (observed in 2014, near solar maximum) decrease to 0.14~keV and 0.7~keV, respectively.

We fitted the spectra of the Perseus cluster beyond   $r=$90$^\prime$  
 from the cluster center and the three observations near the virial radius of the AWM7 cluster.
Here,  we did not use the observations belonging to the N arm due to contamination of a background group \citep{Urban14}.
As shown in figure \ref{fig:bgd}, this background model reproduces the spectra of these regions well.
These spectra show a peak structure around 1 keV, which is a characteristic of $\sim$1 keV plasma.
Figure \ref{fig:08} shows the radial profiles of the emission measure of the HG component(hereafter EM$_{\rm HG}$) beyond   90$^\prime$.  
There are some azimuthal differences: the SE arm shows the brightest EM$_{\rm HG}$, a factor of three higher than that for the NW arm.
Along each arm, EM$_{\rm HG}$ is relatively flat, although
for the S arm, EM$_{\rm HG}$ at 100$^\prime$ and 108$^\prime$ tend to be higher than those beyond 115$^\prime$.
We calculated the weighted average of EM$_{\rm HG}$ for each arm  beyond 95$^\prime$  and plotted them as horizontal lines in figure \ref{fig:08}.
For the S arm, we adopted the weighted average between 90$^\prime$ and 110$^\prime$ since the EM$_{\rm HG}$ is flat in this range but decreases beyond 115$^\prime$.

\subsection{Spectral analysis within the virial radius}
\label{sec:fits}

\begin{figure*}
 \centerline{
\includegraphics[width=4.5cm]{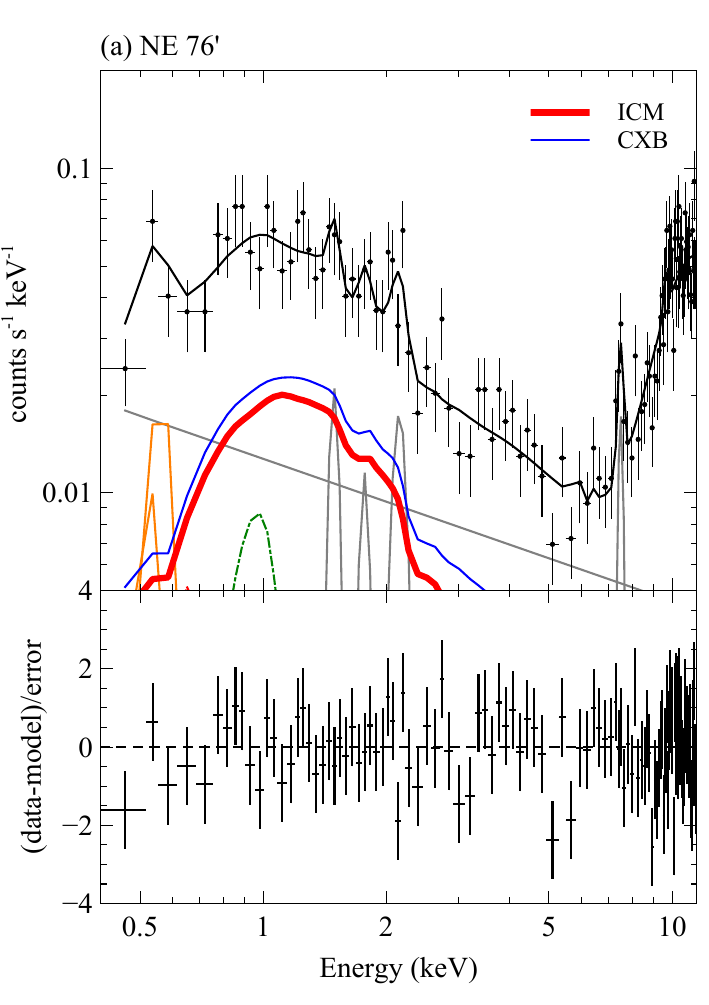}
\includegraphics[width=4.5cm]{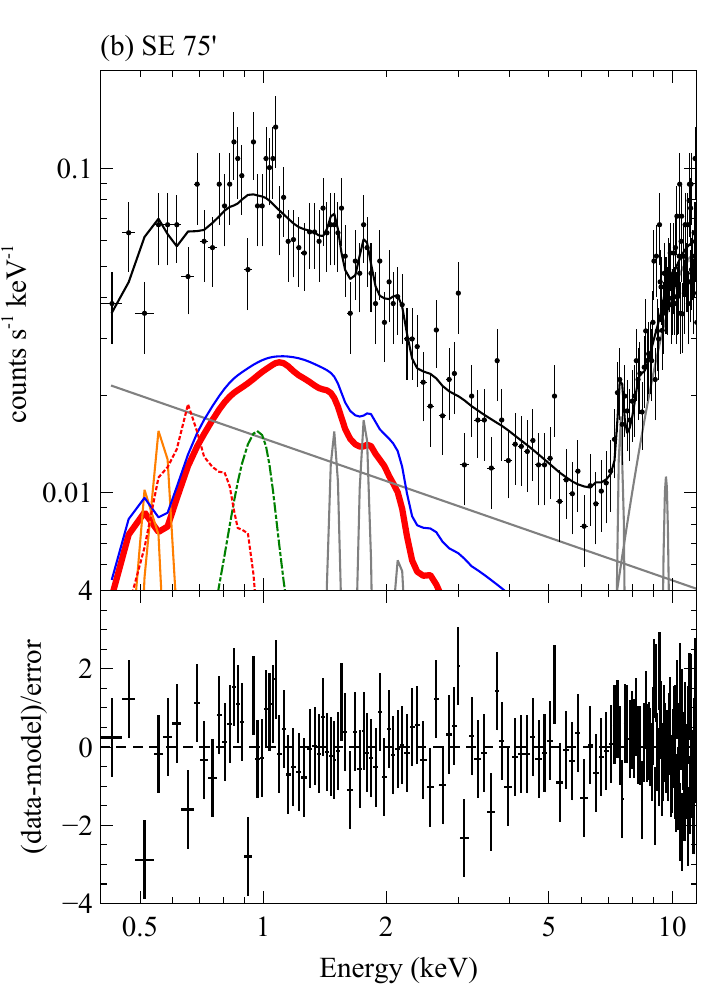}
\includegraphics[width=4.5cm]{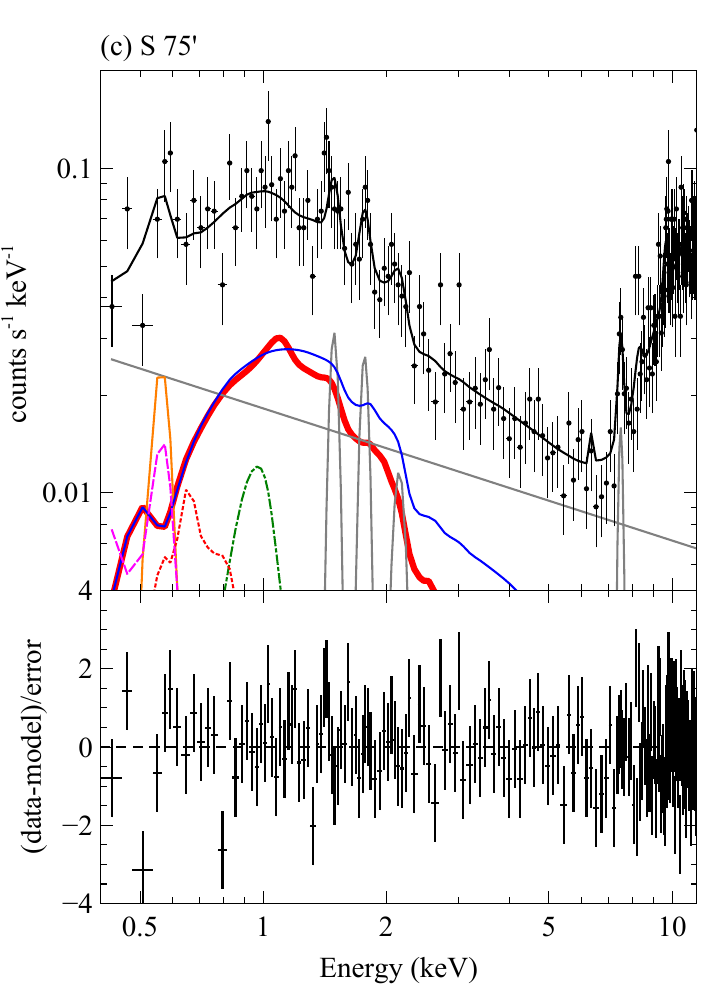}
\includegraphics[width=4.5cm]{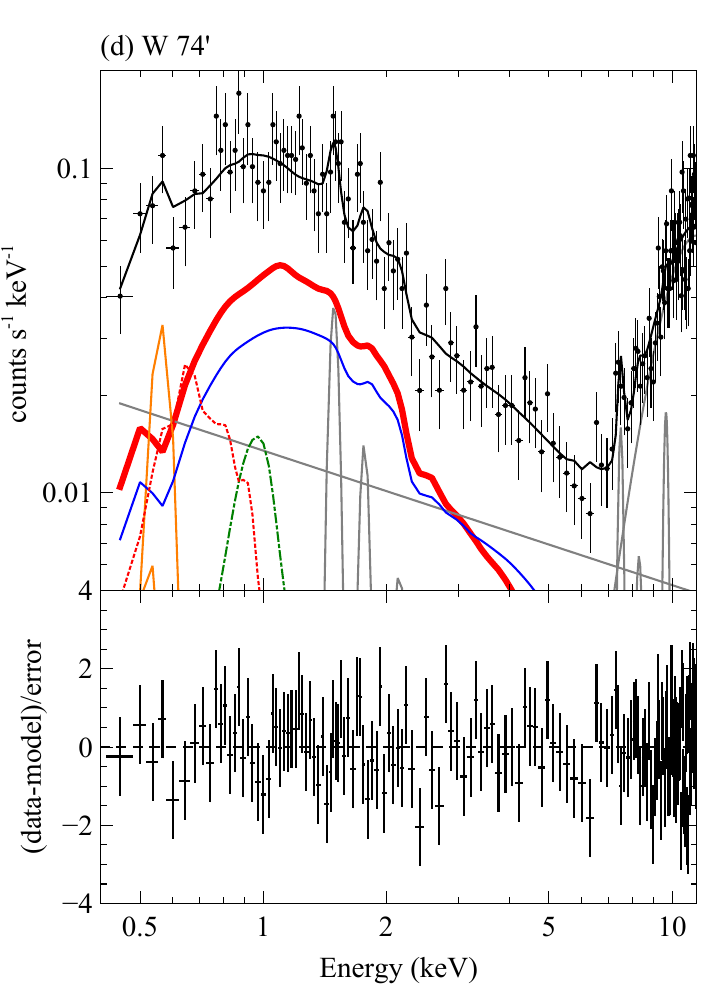}
}
\centerline{
\includegraphics[width=4.5cm]{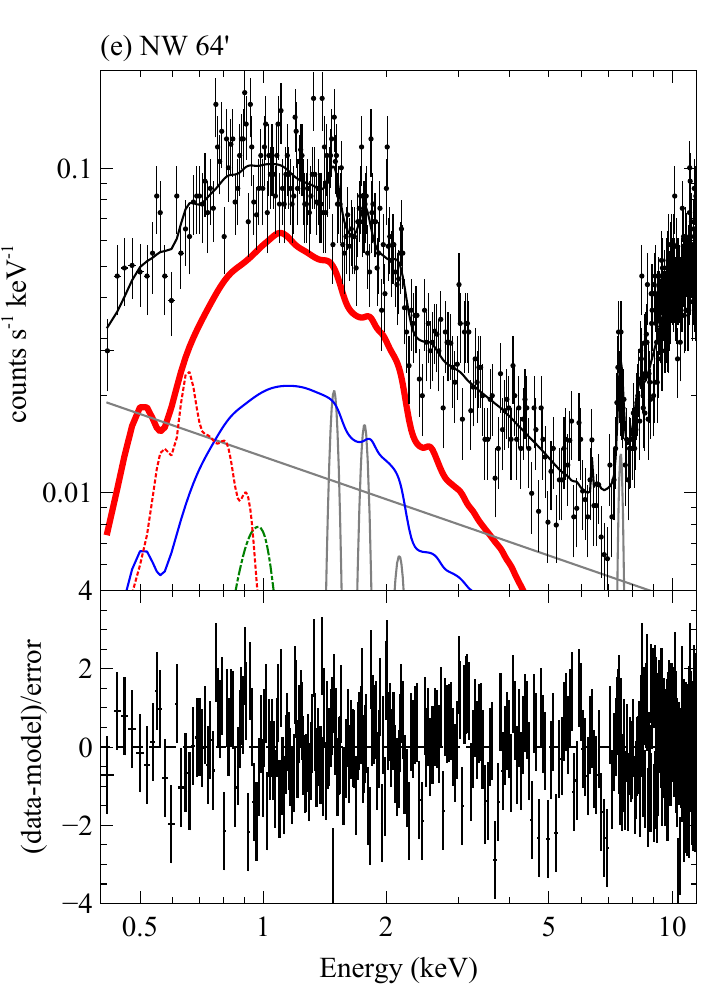}
\includegraphics[width=4.5cm]{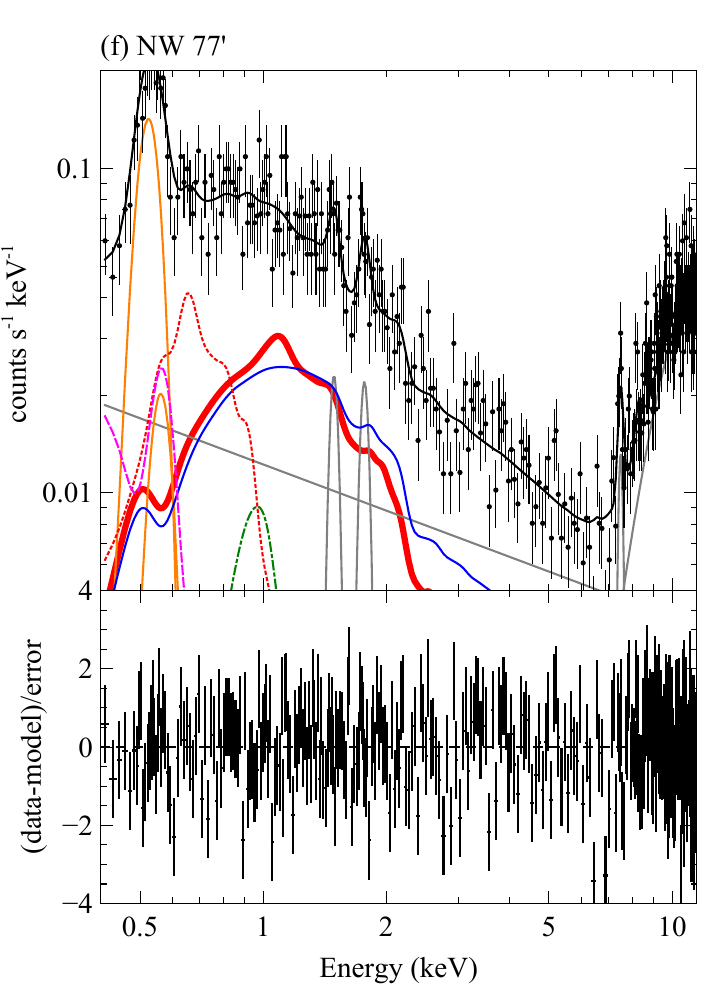}
\includegraphics[width=4.5cm]{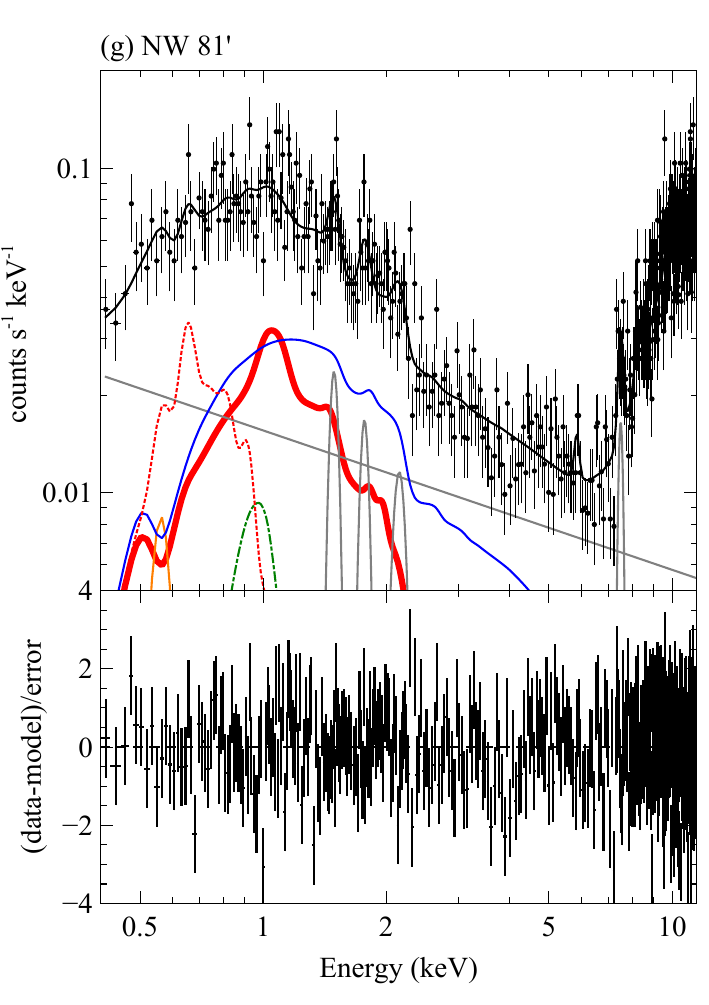}
\includegraphics[width=4.5cm]{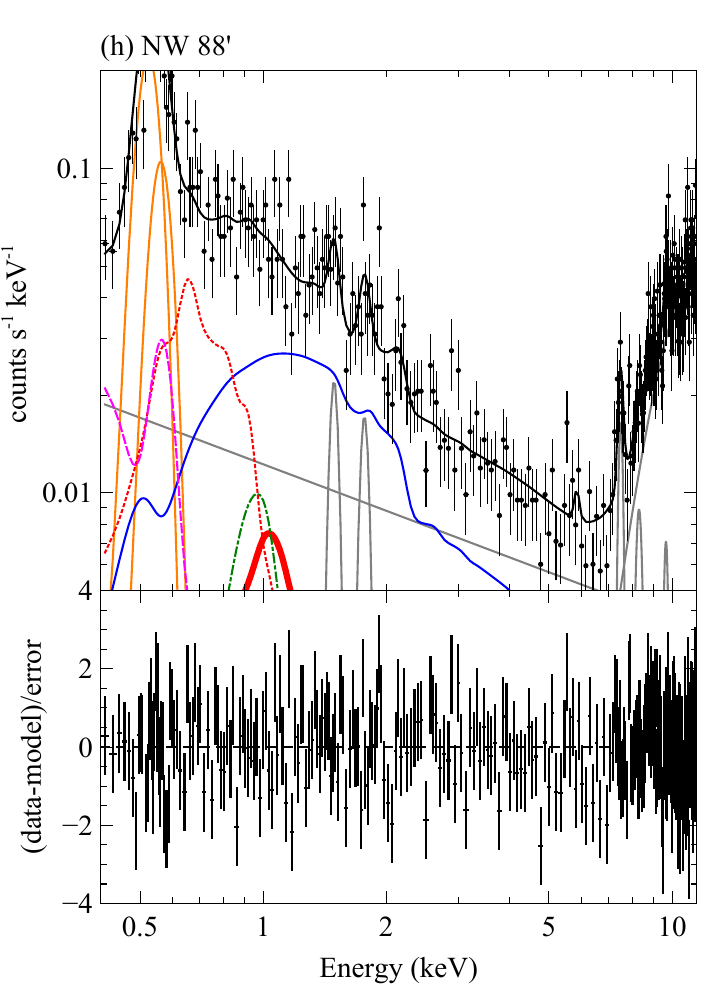}
}

   \caption{The representative XIS1 spectra (upper panels) and residuals (bottom panels). The contributions from the ICM (thick solid red) and the other components (the meaning of colors and line types are the same as in figure \ref{fig:bgd}) are also shown. 
    
{Alt text: Eight subfigures, each consisting of two panels: the upper panels show the representative XIS1 spectra of the Perseus cluster, and the lower panels display the residuals.  Each subfigure corresponds to a different arm direction at radii of approximately 64$^\prime$--88$^\prime$.}  
   }
\label{fig:spec}
\end{figure*}

\begin{figure*}
      \centerline{ \includegraphics[width=13cm]{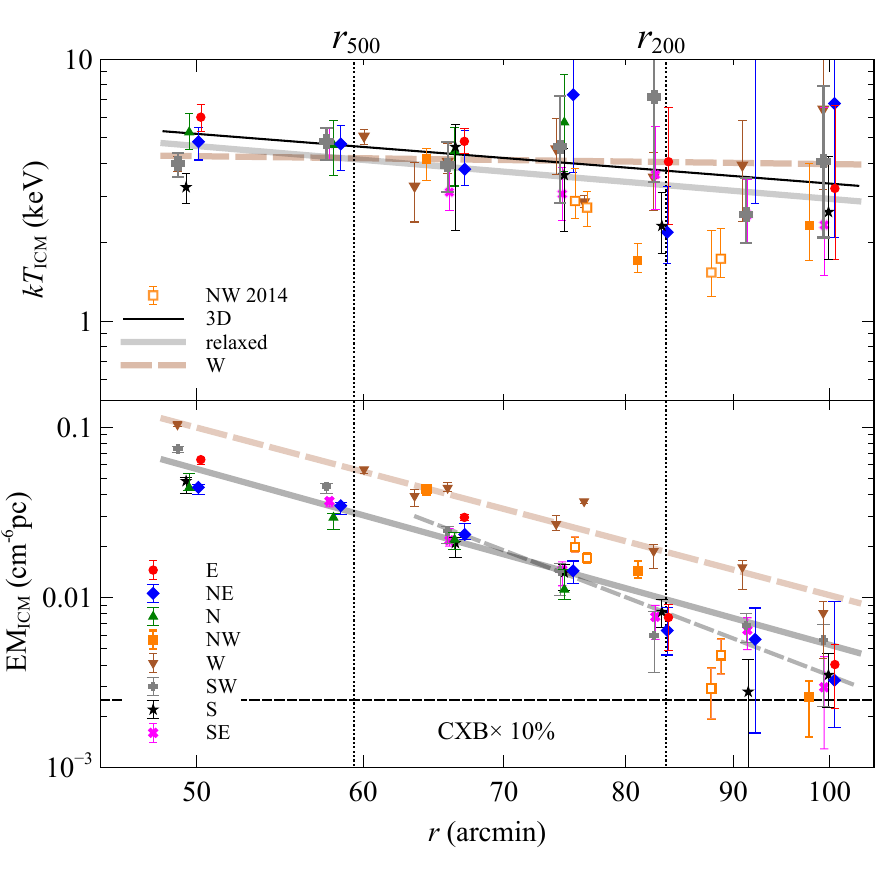} }
\caption{   The radial profiles of $kT_{\rm ICM}$ (upper panel) and EM$_{\rm ICM}$ (bottom panel) of the Perseus cluster 
for the E (the red filled circle), NE (the blue filled diamonds), N (the filled upward triangles), 
NW (the orange-filled squares), W (the brown downward triangles), SW (the gray pluses), S (the black stars), SE (the magenta crosses) arms.
The open squares show the data from the NW arm observed in 2014.
The gray solid and brown dashed lines are the best-fit power-law relations for the relaxed arms (except for the NW and W arms) and the W arm, respectively, over $r=$49$^\prime$--101$^\prime$. 
The black solid line in the upper panel represents the three-dimensional temperature profile (see section 3.5) and
the dashed gray line in the bottom panel shows the best-fit EM$_{\rm ICM}$ profile for $r=63^\prime$--$101^\prime$ radial range.
The two vertical dotted lines correspond to $r_{500}$ and $r_{200}$.
The horizontal dashed line represents a flux level equal to 10\% of the CXB.
{Alt text: Two-panel plot showing radial profiles of ICM temperature (top) and emission measure (bottom) in the Perseus cluster.
Colored symbols represent different arm directions. } }
\label{fig:radial}
\end{figure*}

We then added another CIE ($apec$) component to model the ICM emission   subject to the photoelectric absorption fixed at the Galactic value at each pointing position  
 and fitted the spectra within  $r\sim 100^\prime$.      
Here, the temperature and normalization of the additional component were allowed to vary while the abundance and redshift were fixed at 0.2 solar and 0.018, respectively.
Although \citet{Werner13} found that the abundance of the outskirts of the Perseus cluster is constant at $\sim$ 0.3 solar, fixing the abundance at this value does not change the results.
The LHB and MWH emission measures were allowed to vary, and the EM$_{\rm HG}$ for each arm was fixed at the value plotted as the horizontal line in figure \ref{fig:08}.
Hereafter, we adopt this as our baseline model.
We did not use the pointing at 83$^\prime$ and 92$^\prime$ of the N arm because of the background group \citep{Urban14}.
Figure \ref{fig:spec} shows representative XIS1 spectra.
The HG component makes a significant contribution to the spectral peak around 1 keV.
At $\sim$ 75$^\prime$ from the cluster center, the contribution of the ICM component is comparable to that of the CXB, except for the W arm, which is towards the filaments of the large-scale structure.
At $r=88^\prime$ in the NW arm, the contribution of the ICM component shows a sudden drop compared to those at 77$^\prime$ and 81$^\prime$.

The upper panel of figure \ref{fig:radial} shows the radial profiles of the projected or two-dimensional temperature (hereafter $kT_{\rm ICM}$)  of the ICM component. Beyond $\sim$ $r_{500}$, at a given radius, the scatter in $kT_{\rm ICM}$ is relatively small.  Table \ref{tab:kt} shows the weighted averages of $kT_{\rm ICM}$ with similar distances for the relaxed and W arms.
  Here, we added a systematic error of 0.6 keV to each data to account for the possible spatial variation of the HG component (see section \ref{sec:sys} for details).  
These values are   4--5 keV within 70$^\prime$,  while beyond this radius, they are slightly lower at   2.5--3.5 keV.

The radial profiles of 
 the emission measure (hereafter EM$_{\rm ICM}=\int n_{\rm e}n_{\rm H}ds$, where $n_{\rm e}$, $n_{\rm H}$, and $s$ is the electron density, hydrogen density, and distance along the line of sight, respectively) of the ICM are plotted in the bottom panel of figure \ref{fig:radial}.
The EM$_{\rm ICM}$ of the relaxed arms decreases smoothly with radius with a relatively small scatter. For example,  at $\sim$ 75$^\prime$, the five (NE, N, SW, S, SE) of the relaxed arms show almost the same EM$_{\rm ICM}$ at 0.015 $\rm{cm^{-6}{pc}}$.
In contrast,  for a given radius, EM$_{\rm ICM}$ for the W arm, which is directed toward the west filament from the Perseus cluster, is a factor of 1.5--2 higher than those for the relaxed arms.
The EM$_{\rm ICM}$ toward the NW arm, which is toward the northwest filament from the cluster,  is also as bright as those of the W arm out to 80$^\prime$, while beyond this radius, it suddenly drops, as reported by \citet{Urban14, Zhu2021, Walker2022}.

\begin{table}
            \tbl{The weighted average of $kT_{\rm ICM}$}{
                      \begin{tabular}{cccc}\hline 
 arm &    $r^{*}$ & $N^{+}$ &$kT_{\rm ICM}$ (keV)  \\\hline                
relaxed & 50 & 5 & { 4.4$\pm$0.4} \\ 
relaxed & 58 & 4 &{ 4.8$\pm$0.5}\\ 
relaxed & 66 & 6 & { 4.2$\pm$0.4}\\ 
relaxed & 75 & 5 & { 3.6$\pm$0.7}\\ 
relaxed & 83 & 5 & { 2.6$\pm$0.7}\\ 
relaxed & 92 & 4 & { 2.8$\pm$0.8}\\ 
\hline
W & 49 & 1 & { 3.9$\pm$0.7} \\ 
W & 60 & 1 & { 5.1$\pm$0.7} \\ 
W & 65 & 2 & { 3.7$\pm$0.7}\\ 
W & 74 & 1 & { 4.5$\pm$1.5} \\ 
W & 82 & 1 & { 3.5$\pm$1.1}\\ 
W & 91 & 1 & { 3.9$\pm$2.0}\\ 
\hline
 \hline 
            \end{tabular}}\label{tab:kt}
                \begin{tabnote}
                    \footnotesize
$^*$ The average of  $r$ of the pointing positions.\\
$^+$ The number of pointings\\                    
\end{tabnote}
                \end{table}

\subsection{Systematic uncertainties}
\label{sec:sys}

\begin{figure*}
   \centerline{\includegraphics[width=8cm]{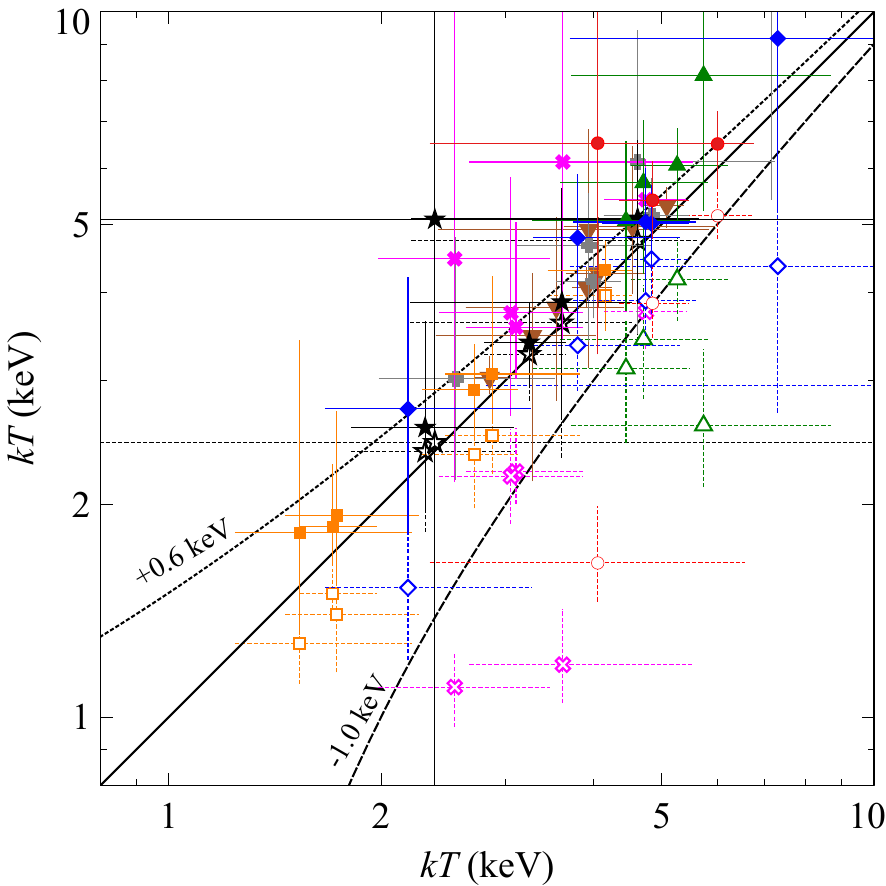}\includegraphics[width=8cm]{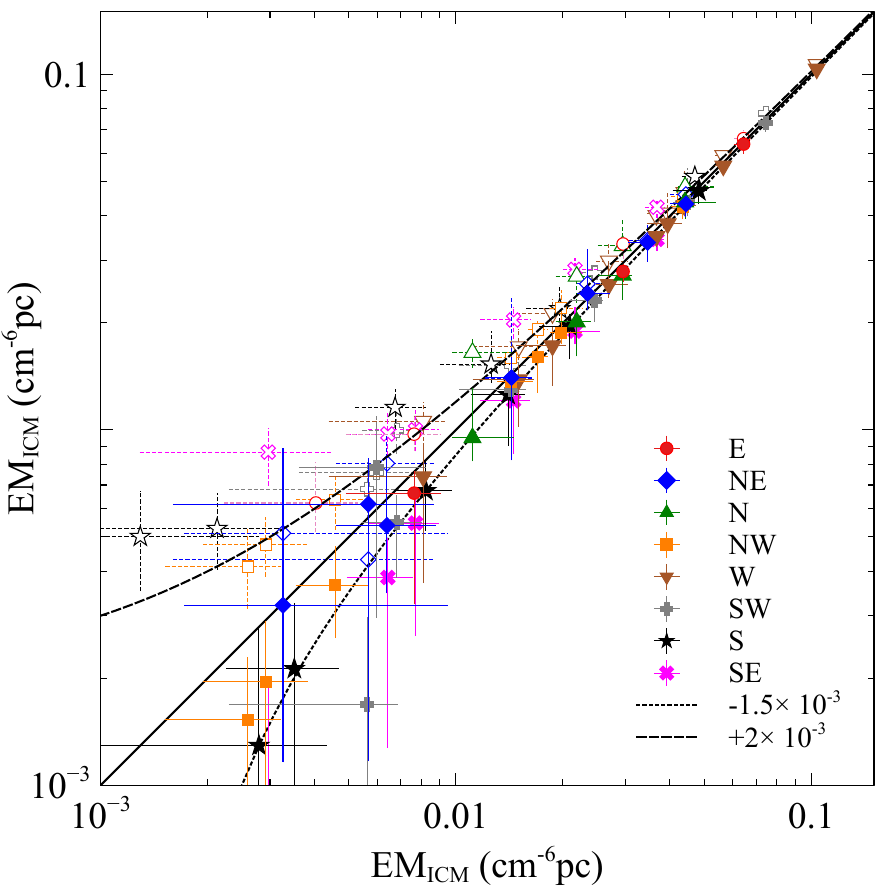}}
\caption{(left panel) $kT_{\rm ICM}$ within 95$'$ with 1.5 times higher HG component (filled symbols with solid error bars) and without the HG component (open symbols with dashed error bars) plotted against those obtained in section \ref{sec:fits}.  The meanings of marks and colors are the same as in figure \ref{fig:radial}. The solid, dotted, and dashed lines correspond to differences of 0.0 keV, + 0.6   keV, and -1.0 keV, respectively. (right panel) The same as the left panel but for EM$_{\rm ICM}$ within 105$'$. The solid, dotted, and dashed lines correspond to differences of 0,  -1.5, and +2.0 , respectively, $\times 10^{-3}{\rm cm^{-6}pc}$.   
{Alt text:  Two scatter plots comparing results from modified background models with the baseline fits for the Perseus cluster.   }}
\label{fig:syshg}
\end{figure*}

Since the EM$_{\rm HG}$ exhibits some azimuthal dependence, it is challenging to constrain their spatial distribution within the virial radius.  
The EM$_{\rm HG}$ varies by a factor of two between the adjacent two arms at most. Signs of spatial variation appear
 along the W, SE, and S arms. Yet, whether this variation arises from the HG component or the ICM emission remains unclear.
To evaluate the effect of the uncertainties in the EM$_{\rm HG}$,
we increased the EM$_{\rm HG}$ by a factor of 1.5 relative to the horizontal lines in figure \ref{fig:radial} and re-fitted the spectra within 100$^\prime$.
Figure \ref{fig:syshg} compares $kT_{\rm ICM}$ and EM$_{\rm ICM}$ obtained from this and the baseline fits. 
Beyond 95$'$, the temperature uncertainties are too large for a meaningful assessment of systematic differences between the models, and these data points were excluded from the plot. 
The typical differences in $kT_{\rm ICM}$ and EM$_{\rm ICM}$ are   0.6 keV and  1.5   $\times 10^{-3}{\rm cm^{-6}pc}$, respectively.

We also fit the spectra with the standard soft X-ray background model, removing the HG component. 
Hereafter, we refer to this model as the standard background model.
The resulting $kT_{\rm ICM}$ and EM$_{\rm ICM}$ are also plotted in figure \ref{fig:syshg}.
  Their radial profiles are shown in figure \ref{fig:sicm} in Appendix \ref{sec:sicm}. 
Without  including the HG component, both $kT_{\rm ICM}$ and EM$_{\rm ICM}$ become flat beyond $\sim 100^\prime$ at $\sim 1$ keV and several $\times 10^{-3} {\rm cm^{-6}pc}$, respectively.
These results support the presence of a relatively flat and slowly varying soft X-ray component with a temperature of $\sim$1 keV, possibly located in the foreground of the Perseus cluster.
The differences in $kT_{\rm ICM}$ and EM$_{\rm ICM}$ between the baseline and the standard background models can reach up to 1 keV and   a few   $\times 10^{-3}{\rm cm^{-6}pc}$, respectively. 
Even around $\sim r_{500}$, the standard background model sometimes yields  $kT_{\rm ICM}$ values 
that are lower  by $\sim 1$ keV.

\begin{figure}
\includegraphics[width=8cm]{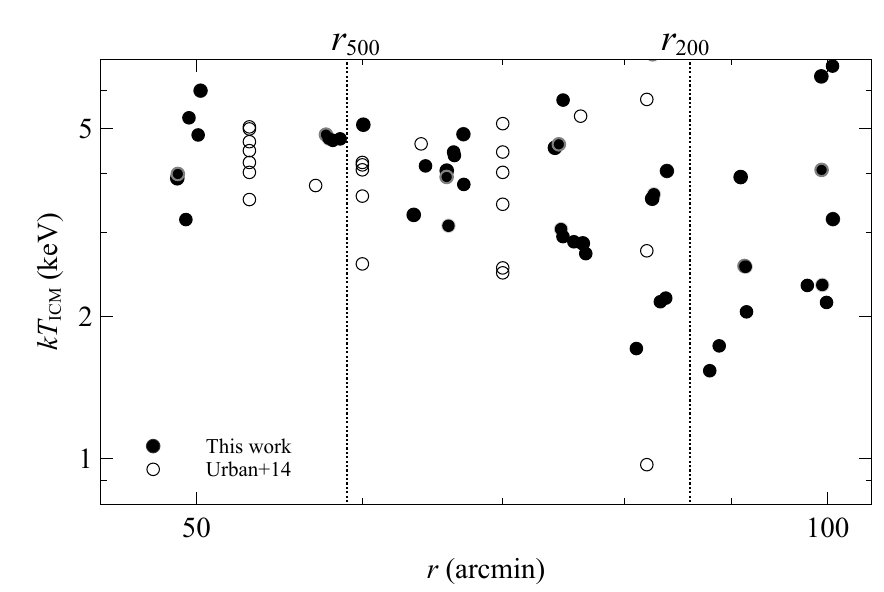}
\caption{The radial profiles of the best-fit $kT_{\rm ICM}$ with this work (filled circles) and by \citet{Urban14} (open circles).
  {Alt text:Two temperature profiles plotted with distinct marker styles.} }
\label{fig:kturban}
 \end{figure}

Figure \ref{fig:kturban} compares our baseline $kT_{\rm ICM}$ profile with that of \citet{Urban14}.
The results are generally consistent, considering differences in the spectral extraction regions.
The data point at $kT \sim 1$ keV around 80$^\prime$ in \citet{Urban14} may be affected by residual HG emission.

 Although most of the data were obtained in 2010-2015 after the solar minimum, the continuous regions of the NW and E arms of the Perseus clusters were observed in 2009, just around the solar minimum. The four overlapping pointings around the virial radius in the NW directions were obtained in 2014 around the solar maximum.
 As shown in figure \ref{fig:spec}, the XIS1 spectrum for the pointing at 76$^\prime$ and 88$^\prime$ along the NW arm shows strong emissions from the O\,\emissiontype {VII} He$\alpha$ line at 0.56 keV and the O\,\emissiontype {I} line at 0.525 keV.
 The former and latter possibly come from the heliospheric SWCX and scattered photons from the  Earth's atmosphere, respectively (appendix \ref{soft}), although at the CCD energy resolution, the normalizations of the O\,\emissiontype {VII} He$\alpha$ line and LHB are strongly coupled.
 However,  $kT_{\rm ICM}$ and EM$_{\rm ICM}$ of the NW arm obtained in 2009 and 2014 are continuous with radius (figure \ref{fig:radial}). 
  Since \citet{Paper I} reported that the O\,\emissiontype {VIII} Ly$\alpha$ line, possibly from SWCX, also contaminates the spectra at the solar maximum, we tried to add this line to the spectra obtained in 2014. However,  $kT_{\rm ICM}$  and  EM$_{\rm ICM}$ were not changed within the statistical error bars.
 Some X-rays are reflected on light paths other than the nominal reflection of the X-ray telescopes (XRT) onboard {\it Suzaku}
 \citep{Mori2005, SuzakuXRT, Takei2012}.  
These X-rays are called stray light and can contaminate emissions near bright sources.
 The stray light contribution depends strongly on the azimuthal angle of the cluster center and the roll angle of the FOV  \citep{Takei2012, Urban14} .
The observations of the NW arm in 2009 and 2014 were obtained with different roll angles,
and the stray light contamination is expected to be quite different.
However, 
 at a given distance from the cluster center, $kT_{\rm ICM}$ and EM$_{\rm ICM}$  obtained in 2009 and 2014 are consistent. 

\begin{figure}
   \centerline{\includegraphics[width=4cm]{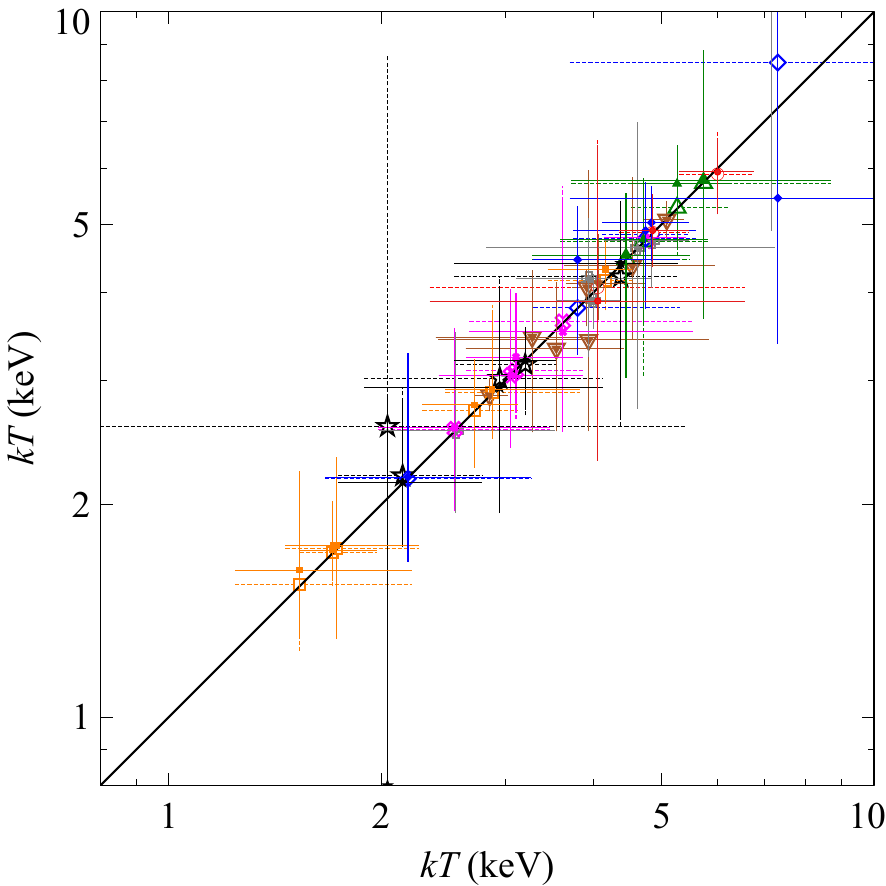}\includegraphics[width=4cm]{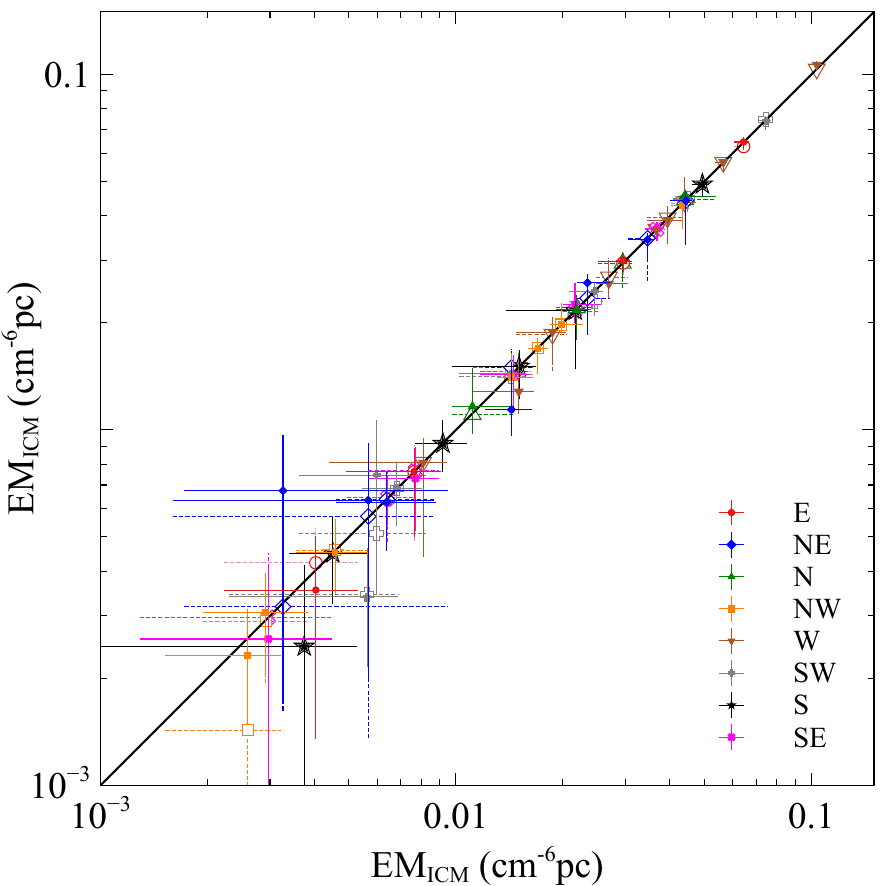}}
   \caption{
(left) The ICM temperature, $kT_{\rm ICM}$ within 95$'$ obtained from spectra without binning (open symbols with dashed error bars) and those with optimal binning (filled symbols with solid error bars), plotted against the values obtained in Section~\ref{sec:fits}.
The meanings of the symbols and colors are the same as in Figure~\ref{fig:radial}.
The solid line indicates equality.
(right) The same as the left panel, but for the emission measure, EM$_{\rm ICM}$ within 105$'$.
Alt text: Two scatter plots comparing the ICM temperature (left) and emission measure (right) derived using different binning schemes with those from the baseline fits for the Perseus cluster.
}
\label{fig:sysbin}
\end{figure}

Although we grouped the spectra to have at least one count per bin, in principle, grouping is not required when using the C-statistic.
In our analysis, the NXB is not subtracted; instead, it is explicitly modeled as an additive spectral component. Therefore, the fits are not affected by issues related to background subtraction.
To verify the robustness of our results, we repeated the spectral fitting without any grouping.
In addition, \citet{Kaastra2016} introduced the concept of optimal energy grids, i.e., optimal binning.
We also fitted the spectra using the optimal binning method implemented in the \textit{ftgrouppha} tool in HEASoft.
As shown in Figure~\ref{fig:sysbin}, the resulting temperatures and normalizations differ only marginally; the differences are much smaller than the corresponding statistical uncertainties.
We therefore conclude that our binning methods do not bias our results, and that our conclusions are insensitive to the choice of binning.

\subsection{Modeling of the emission measure profiles}
\label{sec:em}
 \begin{table*}
            \tbl{The power-law modeling for $kT_{\rm ICM}$ and EM$_{\rm ICM}$}{
                      \begin{tabular}{cccccc}\hline 
 arm  &   $r$   &  &  $kT_{\rm 59}^*$ or EM$_{\rm 59}^*$ & index$^+$    & $\chi^2$/d.o.f                   \\
  &   (arcmin)  & & (keV) or ($\rm{cm^{-6}pc}$) \\ \hline
relaxed & 49$^\prime$--{ 101}$^\prime$ & EM$_{\rm ICM}$  & { $(3.22\pm 0.06)\times 10^{-2}$} & { 3.43$\pm 0.11$} & {  128/32}\\
relaxed & { 63$^\prime$}--{ 101}$^\prime$ & EM$_{\rm ICM}$  &{  $(4.25\pm 0.32)\times 10^{-2}$} & { 4.76$\pm 0.38$} & { 20/21}\\
relaxed  & 49$^\prime$--{ 101}$^\prime$ & $kT_{\rm ICM}$ &  { 4.17$\pm$0.22}  & { 0.67$\pm$0.25} & { 20/32 }\\ 
\hline
W  & 49$^\prime$--{ 101$^\prime$} &EM$_{\rm ICM}$  & ${ (5.79\pm 0.14)\times 10^{-2}}$ & { 3.28$\pm 0.15$} & {  5/6}\\
W   & 49$^\prime$--{ 101$^\prime$} & $kT_{\rm ICM}$ & {  4.19$\pm$0.36}  & { 0.1$\pm$0.5} & { 3/6} \\ 
\hline
            \end{tabular}}\label{tab:powerlaw}
                \begin{tabnote}
                $^*$  $kT_{\rm ICM}$ or EM$_{\rm ICM}$ at 59$^\prime$.\\
                $^+$  The index for the power-law modeling. \\
                    \footnotesize
\end{tabnote}
                \end{table*}

We model the radial profiles of EM$_{\rm ICM}$ using a power-law formula as follows,

\begin{equation}
 \rm{EM}_{\rm ICM}={\rm EM}_{\rm 59} {\it \left(\frac{r}{\rm 59^\prime}\right)^{-b}}
\end{equation}
where,  EM$_{59}$ is EM$_{\rm ICM}$ at $r=59^\prime$ and $b$ is the power-law index.
Here,  a systematic error of  1.5$\times 10^{-3}~ {\rm cm^{-6}pc}$   is added to account for the possible spatial variation of the HG component.
The fitting results are summarized in table \ref{tab:powerlaw}.
Fitting the relaxed arms over \(r=49^\prime\text{--}101^\prime\) yields $b=3.43\pm 0.11$, but provides a poor description $\chi^2=128$ for 32 degrees of freedom (d.o.f), and \(\mathrm{EM}_{\rm ICM}\) tends to fall below the best-fit model beyond $80^\prime$.
The large $\chi^2$ is driven in part by enhanced azimuthal scatter in surface brightness around $r\sim 50^\prime$, where residual substructure associated with the large-scale sloshing/swirling pattern reported for Perseus \citep{Simionescu12, Churazov2025} likely remains.
 To mitigate this, we restrict the fit to $r=63^\prime\text{--}101^\prime$, obtaining a steeper slope $b = 4.8 \pm 0.4$ with an acceptable goodness of fit $(\chi^2/\mathrm{d.o.f.}=20/21)$. As shown in Figure~\ref{fig:radial}, this power-law model reproduces \(\mathrm{EM}_{\rm ICM}\) well over the adopted range. 
Fitting the W arm over \(r=49^\prime\text{--}101^\prime\) yields a slope similar to that of the relaxed arms, but an  \(\mathrm{EM}_{59}\) higher by a factor of 1.8.  Here, we exclude the data point at \(r=76^\prime\) south of the main W arm.

\begin{figure}
      \centerline{ \includegraphics[width=8cm]{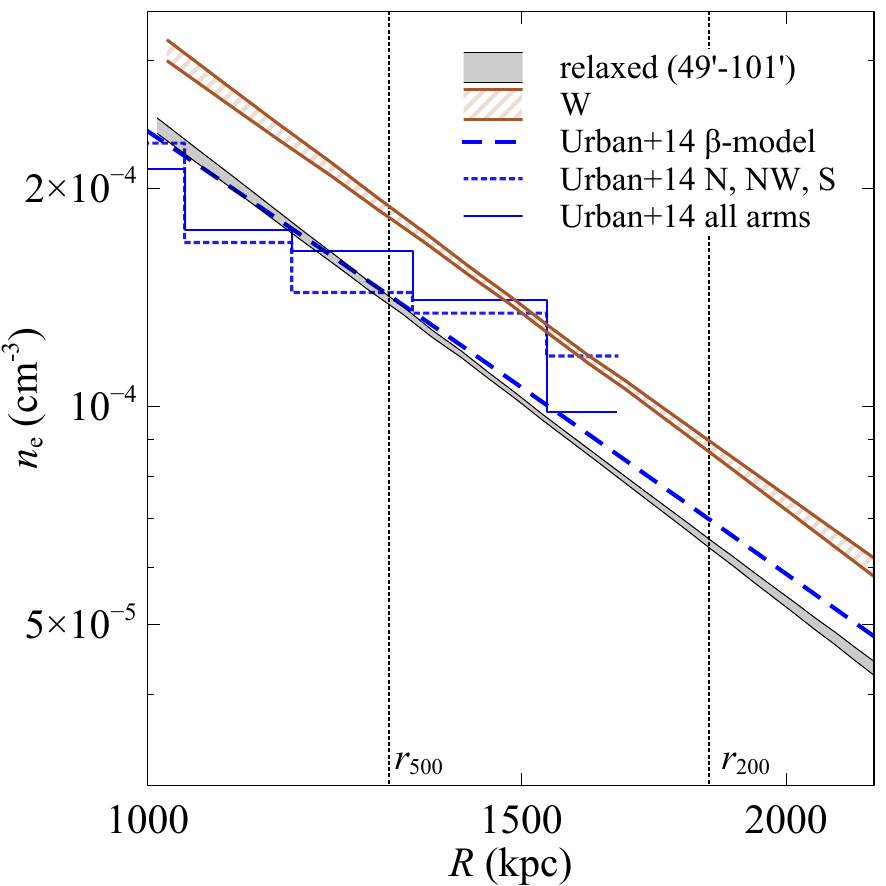} }
\caption{ The radial profiles of $n_{\rm e}$ of the Perseus cluster for the relaxed arms 
(the gray filled area for 1 $\sigma$ error range at   $r$=49$^\prime$--101$^\prime$)   and for the W arm (brown hatched area for 1 $\sigma$ error range).
 The dashed blue line is the best-fit $\beta$ model beyond 10$^\prime$ of all arms by \citet{Urban14}. The blue solid and dotted lines show $n_{\rm e}$ by \citet{Urban14} for all and N, NW, and S arms, respectively.
   {Alt text:A line graph showing electron density profiles versus radius in kiloparsecs. Multiple curves are plotted: a gray shaded band, a brown hatched region, and three blue curves with different line styles.} }
\label{fig:radialne}
\end{figure}

When EM$_{\rm ICM}$ is modeled by a power-law, assuming spherical symmetry, the three-dimensional density profile is expressed by,
\begin{equation}
 {n_e}={n_e}_{59} \left(\frac{R}{1.3 ~\rm Mpc}\right)^{-(1+b)/2}	
\end{equation}
where, $n_e$ and ${n_e}_{59}$ are the electron density and that at $r=59^\prime$ or 1.3 Mpc, respectively, and $R$ is the three-dimensional distance from the cluster center.
The radial profiles of $n_{\rm e}$ are plotted in figure \ref{fig:radialne}.
The results are summarized in table \ref{tab:powerlaw2}.
As shown in figure \ref{fig:radialne}, this best-fit density profile for the relaxed arm shows excellent agreement with the best-fit $\beta$-model profile beyond 10$^\prime$ of the Perseus cluster by \citet{Urban14}, whose density profiles deviate from the $\beta$-model beyond $r_{500}$.
Note that the definition of "relaxed arms" by \citet{Urban14} is "N, NW, S" arms and differs from ours.
Using the data beyond $r=49^\prime$, the relaxed and the W arms have the same slope at  2.1--2.2, but the W arm has a factor of 1.3 higher $n_e$ at a given radius.
This slope is consistent with the value measured for the X-COP clusters,  $2.47\pm0.31$, over 1.15--2.00 $r_{500}$ by \citet{XCOP2019}.
Restricting the fit for the relaxed arms to $r=63^\prime$--$101^\prime$  yields a steeper slope, $2.9\pm0.2$.
At $r_{500}$ (1.3 Mpc), we find $n_e=1.4\times10^{-4}~\mathrm{cm^{-3}}$ for the relaxed arms, in agreement with the X-COP median.
At $r_{200}$ (1.8 Mpc), the relaxed-arm fits over $49^\prime$--$101^\prime$ and $63^\prime$--$101^\prime$ give $n_e=6\times10^{-5}~\mathrm{cm^{-3}}$ and $5\times10^{-5}~\mathrm{cm^{-3}}$, respectively, both close to the X-COP median of $5\times10^{-5}~\mathrm{cm^{-3}}$.
Here, $r_{500}$ and $r_{200}$ are derived under the assumption of hydrostatic equilibrium (Section~\ref{sec:hm}).

\subsection{Modeling of temperature profiles}

We apply a power-law modeling of $kT_{\rm ICM}$ as follows,

\begin{equation}
 {kT}_{\rm ICM}={kT}_{59} \left(\frac{r}{\rm 59^\prime}\right)^{-a}
\end{equation}
where, $kT_{59}$ is  $kT_{\rm ICM}$ at $r=59^\prime$, 
and $a$ is the power-law index.
A systematic error of 0.6 keV is included in each data to account for the possible spatial variation of the HG component.
As shown in figure \ref{fig:radial}, the radial profiles of $kT_{\rm ICM}$ are consistent with this power-law model.
For the relaxed arms, using the data within  $r=49^\prime$--101$^\prime$ , we obtained $kT_{59}= 4.17 \pm 0.22$ keV and  $a=0.67\pm 0.25$ ,  as summarized in table \ref{tab:powerlaw}. 
The power-law fit to $kT_{\rm ICM}$ for the W arm, using the data within $r=49^\prime$-- 101 $^\prime$, gives us values $kT_{59}$ and $a$ consistent within error bars with those for the relaxed arms. 
Here, we excluded the data point at $r=76^\prime$ south of the main W arm, whose $kT_{\rm ICM}$ is lower than the others.

We then model a three-dimensional temperature profile, $ {kT}_{\rm 3D,  ICM}$,  assuming the same power-law index, as follows, 

\begin{equation}
 {kT}_{\rm 3D,  ICM}={kT_{\rm 3D, 59}} \left(\frac{r}{\rm 59^\prime}\right)^{-a}
\end{equation}

where $kT_{\rm 3D, 59}$ is  $kT_{\rm ICM}$ at $r=59^\prime$.
Using the three-dimensional density profile calculated in section \ref{sec:em},  
we created mock spectra considering the projection effect and fitted them. If we adopt ${kT_{\rm 3D, 59}}$=4.7 keV,   
we can recover
the projected temperature profile within a few percent. The best-fit 3D profile is plotted in figure \ref{fig:radial}.

 \begin{table*}
            \tbl{The power-law indices for the radial profiles of $n_{\rm e}$, $K$, and $P$, and $r_{500}$ and $r_{200}$}{
                      \begin{tabular}{cccccccc}\hline 
   arms  &   &  index for $n_{\rm e}$ &  index for $P$ &   index for $K$      &$r_{500}$/$r_{200}$      \\
\hline
             relaxed  & 49$^\prime$-{ 101}$^\prime$ &   { 2.21$\pm$0.06}  &  {  2.88$\pm$0.26} &  { 0.81$\pm$0.25} & 1{ .3/1.8 Mpc (59$^\prime$/84$^\prime$)}  \\
   relaxed  & 63$^\prime$-{ 101}$^\prime$ & { 2.88$\pm$0.19} & { 3.6$\pm$0.3} & { 1.3$\pm$0.3} & { 1.4/2.0 Mpc (64$^\prime$/90$^\prime$) }\\
          W & 49$^\prime$-{ 101}$^\prime$ & { 2.14$\pm$0.07}  & {   2.2$\pm$0.5} &  { 1.3$\pm$0.5} & { 1.1/1.7 Mpc (50$^\prime$/78$^\prime$)}  \\ \hline
            \end{tabular}}\label{tab:powerlaw2}
                \begin{tabnote}
                    \footnotesize
\end{tabnote}
                \end{table*}


 %
	
%

\section{Discussion}

\subsection{The hydrostatic mass}
\label{sec:hm}

\begin{figure}
\includegraphics[width=8cm]{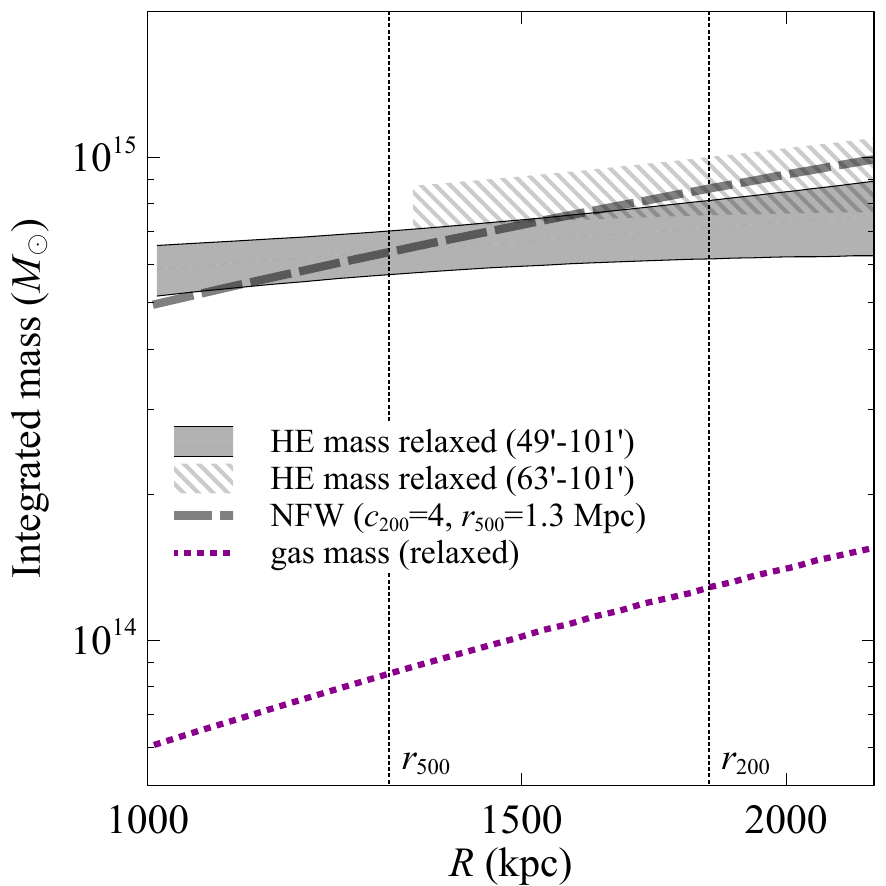}
\caption{Radial profiles of the hydrostatic mass 
    of the relaxed arms using the power-law relations for the EM$_{\rm ICM}$ 
 of 49$^\prime$--$101^\prime$  (gray filled area) and 63$^\prime$--101$^\prime$ (hatched area).    
  The solid and dashed lines represent the NFW mass models for $c_{200}=4$ with $r_{500}= 1.30$ Mpc. 
    The dotted line shows the integrated gas mass profile for the relaxed arms ($r=49^\prime$--$101^\prime$).
     {Alt text:Mass profiles with shaded and hatched regions, overlaid with solid, dashed, and dotted lines.} } 
\label{fig:mass}
\end{figure}

From the best-fit power-law relations for $kT_{\rm ICM}$ and EM$_{\rm ICM}$   of the relaxed arms , we estimated the total gravitational mass assuming that the ICM is in hydrostatic equilibrium and spherical symmetry (hereafter referred to as the hydrostatic mass).
The resultant mass profile is plotted in figure \ref{fig:mass} and the derived values of   $r_{500}$ and $r_{200}$   are summarized in table \ref{tab:powerlaw2}.
Here, we adopted the   three-dimensional  $kT_{\rm 3D, ICM}$ profile.  
Using the data for the relaxed arms of  $r=49^\prime$--101$^\prime$, 
we obtained $r_{500}=59^\prime$ or 1.3 Mpc, $M_{500}=6.4\times 10^{14} M_\odot$, and 
 $r_{200}$=84$^\prime$ or 1.8 Mpc.  Unless otherwise stated, we adopt these values as our reference $r_{500}$ and $r_{200}$.  
This value for $r_{500}$ is consistent with that derived from the pressure profile of the Perseus cluster by \citet{Urban14}. 
Figure \ref{fig:mass} also shows a Navarro-Frenk-White (NFW) mass model \citep{NFW} with a concentration parameter, $c_{200}=4$.
The slope of the hydrostatic mass is slightly smaller than that of the NFW mass model.
If we adopt the slope of EM$_{\rm ICM}$ of 63$^\prime$--101$^\prime$, we get a higher hydrostatic mass by $\sim$20\%, and it becomes more consistent with the NFW mass around $r_{200}$.
From $\sim r_{500}$ to $\sim r_{200}$, the scatter in $kT_{\rm ICM}$ and EM$_{\rm ICM}$ at a given radius is quite small.
Excluding the W and NW arms, which are towards the filaments,  the ICM in the Perseus cluster outskirts seems regular and relaxed in near hydrostatic equilibrium.

\subsection{The gas and baryon fraction}
\label{sec:fb}

Since the dark matter and baryons accrete from the surroundings, the baryon fraction at the cluster outskirts is expected to match the cosmic baryon fraction.
At $r_{500}$, the integrated baryon fraction (stellar fraction + gas fraction) agrees with the cosmic baryon fraction for massive clusters with $M_{500}$ higher than several times 10$^{14} M_\odot$ \citep{Akino2022}.
Beyond $r_{500}$, based on the Suzaku observations of the Perseus cluster, \citet{Simionescu11} and \citet{Urban14} reported that the gas fraction exceeds the cosmic value and discussed the gas clumping causes an overestimation of gas density. In contrast, the ratios of the integrated gas mass with Suzaku to weak lensing mass with Subaru do not exceed the cosmic fraction \citep{Kawaharada10, Ichikawa13, Okabe14b}.

\begin{figure}
\includegraphics[width=8cm]{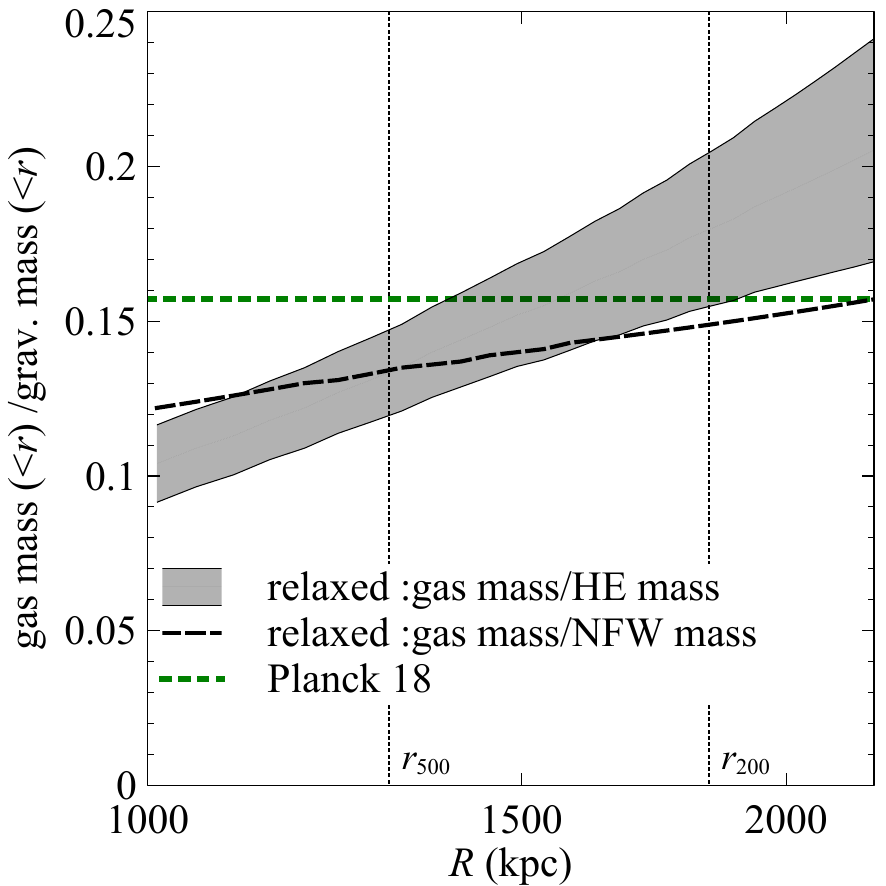}
\caption{Radial profiles of the gas fraction, the ratio of the integrated gas mass to the hydrostatic mass for the relaxed arms using the power-law density profile for the   $r=49^\prime$--$101^\prime$   data (the gray filled area),  and that to the NFW mass with $c_{200}=4$ and $r_{500}=1.30$ Mpc (the dashed line).
The dotted horizontal line shows the cosmic baryon fraction by \citet{Planck18}. 
   {Alt text: A plot of the gas fraction versus radius in units of kpc. } }
\label{fig:fgas}
\end{figure}

Figure \ref{fig:mass} also shows the integrated gas mass, calculated using the $\beta$-model by \citet{Urban14} within 0.9 $r_{500}$, or $53^\prime$, and beyond that radius, using the best-fit power-law density profile for the relaxed arms at   $r=49^\prime$--101$^\prime$.  
The radial profile of the integrated gas fraction, defined as the ratio of the integrated gas mass to the hydrostatic mass for the relaxed arms, is shown in Figure \ref{fig:fgas}.
For the relaxed arms,  the integrated gas mass fractions are 0.13$\pm$0.01 and 0.18$\pm$0.02 at $r_{500}$ and $r_{200}$, respectively, when using the power-law relation for 49$^\prime$--$101^\prime$. Since using the power-law relation for 63$^\prime$--101$^\prime$ yields an integrated mass profile more consistent with the NFW mass, we also derived the gas fraction with the NFW model, adopting $r_{\rm 500}=1.30$ Mpc and $c_{200}=4$, obtaining the gas fractions of 0.13 at $r_{500}$ and 0.15 at $r_{200}$.  
 With the two-micron all-sky survey (2MASS) data, the total K-band stellar luminosity
out to $r_{200}$ of the Perseus cluster has been calculated by  \citet{Matsushita2013}.
 Assuming that the mass-to-light ratio of the stellar mass is around unity \citep{Nagino2009}, the ratio of the stellar mass to the hydrostatic mass within  $r_{200}$ is 0.013.  Then, adding the stellar mass fraction, the baryon fraction for the relaxed agrees with the cosmic baryon fraction of 0.157 from the Planck 2018 results \citep{Planck18}, as expected.
We do not need gas clumping, at least for the relaxed arms.

\subsection{The pressure profiles}

\begin{figure}
\centerline{\includegraphics[width=8cm]{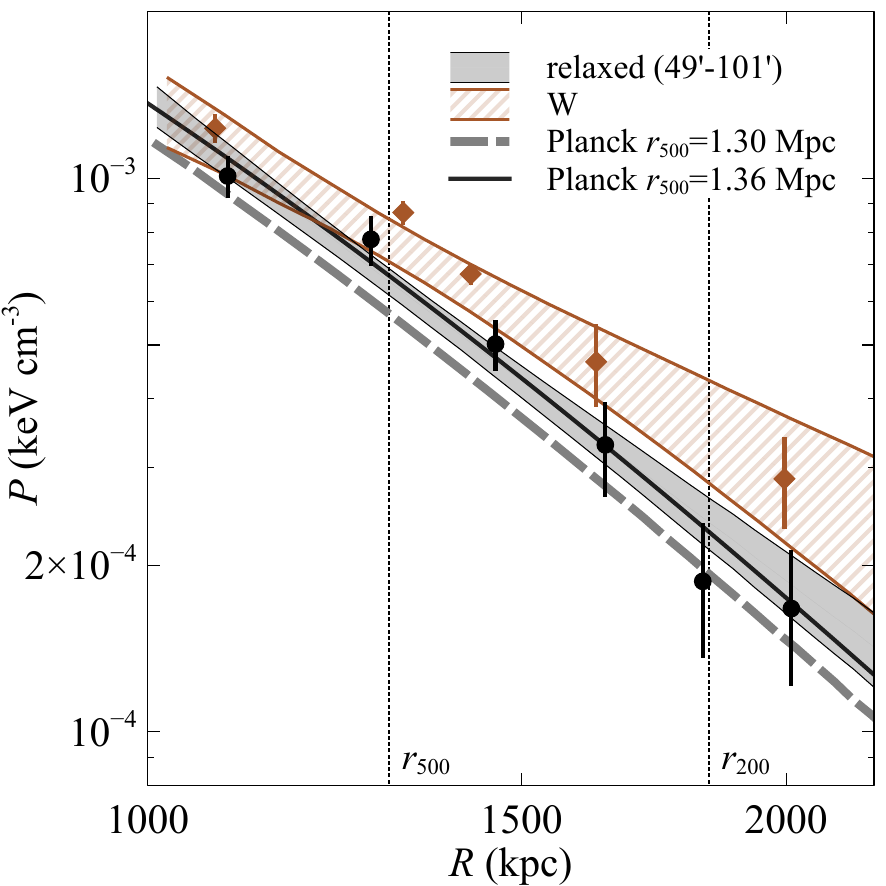}}
\caption{The radial profiles of thermal electron pressure.
The meaning of lines and colors is the same as in figure \ref{fig:radialne}.
The data points (filled circles for the relaxed arms and filled diamonds for the W arm) are from the weighted averages of $kT_{\rm ICM}$ shown in table \ref{tab:kt} and the power-law fits for EM$_{\rm ICM}$.
The gray dashed and black solid lines show the average profiles with Planck collaboration (2013) for $r_{500}=1.30$ Mpc and 1.36 Mpc, respectively.
   {Alt text:A plot of thermal electron pressure versus radius. Multiple curves with different line styles and two sets of data points (circles and diamonds) are overlaid.}} 
\label{fig:p}
\end{figure}

A generalized pressure profile for the ICM is proposed by
\citet{Nagai2007}.  \citet{Arnaud2010} found that this profile well represents the ICM observed with XMM-Newton.
The thermal Sunyaev-Zeldvich (SZ) effect, caused by the inverse Compton
scattering of the cosmic microwave background photons by the ICM,
is proportional to the thermal gas pressure integrated along the line of sight.
The SZ measurements are complementary to the X-ray observations,
which measure the temperature and emission measure of the ICM.
With the Planck satellite, \citet{Planck2013} found that their pressure profiles
agree well with the XMM-Newton results of \citet{Arnaud2010} within $r_{500}$.
We calculated the thermal electron pressure profiles using the power-law model fits summarized in table \ref{tab:powerlaw}.
The power-law indices of the pressure are    2.88$\pm$0.26 for the relaxed arms using the $r=49^\prime$--101$^\prime$ (table \ref{tab:powerlaw2}).   
In addition, we also used the weighted averages of $kT_{\rm ICM}$ (table \ref{tab:kt}), 
 multiplied a factor of $\sim$ 1.1 to recover the three-dimensional values,  
 and the power-law density profiles to calculate the pressure profiles.  
As shown in figure \ref{fig:p}, these two methods give consistent pressure profiles.
Along the relaxed arms, the ICM pressure profile is slightly higher than the average profile with Planck for a 
$r_{500}=1.30$ Mpc cluster \citep{Planck2013}.
Fitting the obtained pressure profile 
with the Planck pressure profile gives a slightly larger $r_{500}$ of 1.36 Mpc.
In contrast,  the pressure towards the W arm is significantly higher than that for the relaxed arms.
Since the $kT_{\rm ICM}$ profile for the W arm is consistent with that for the relaxed arms, the difference in the pressure profile is caused by the higher ICM density toward the W arm.

\subsection{The entropy profiles}

The entropy, $K$, is given by

\begin{equation}
	K= kT n_e^{-2/3}
\end{equation}

\begin{figure}
\centerline{\includegraphics[width=8cm]{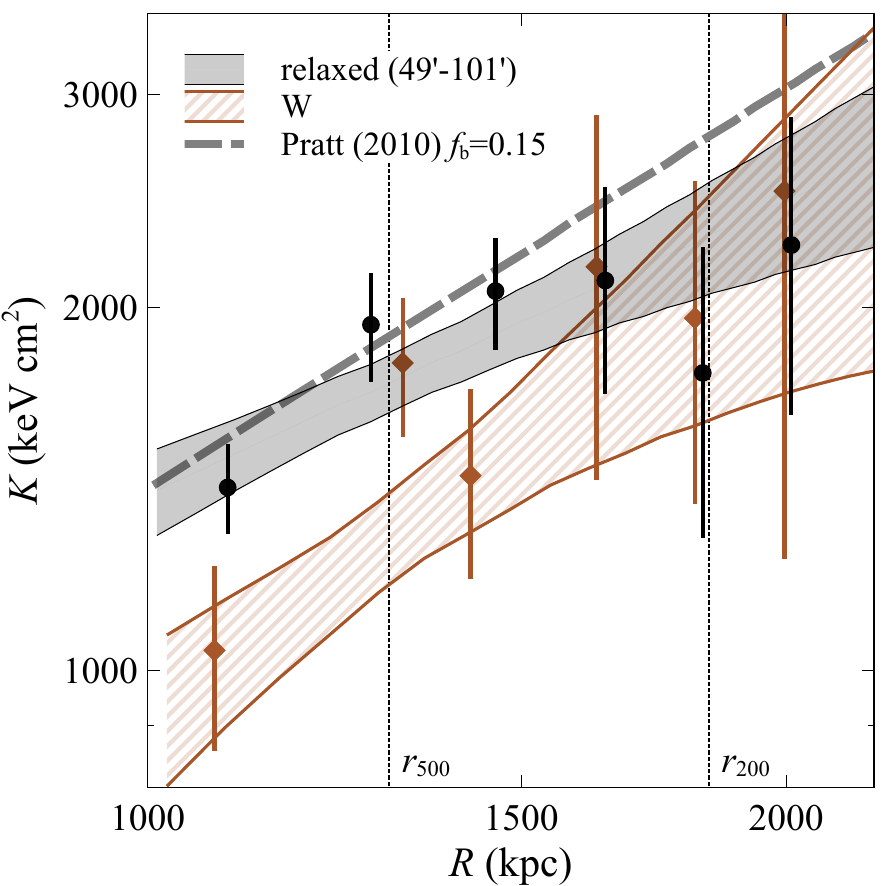}}
\caption{Radial profiles of entropy. The meaning of lines and colors is the same as in figure \ref{fig:radialne}.  
The data points (filled circles for the relaxed and filled diamonds for the W arms) are derived from the average temperatures shown in table \ref{tab:kt}.
The gray dashed line shows 
an $R^{1.1}$ power-law model with normalization fixed to the expected value calculated by Pratt et al. (2010) for $r_{500}$=1.30 Mpc.
   {Alt text: A plot of entropy versus radius. Multiple curves with different line styles and two sets of data points (circles and diamonds) are overlaid.} }
\label{fig:K}
\end{figure}

As clusters grow, shocks heat the accreting gas, and $K$ is expected to increase with radius.
From numerical simulations of gravitational collapse,
 the entropy is expected to follow the following form \citep{Voit2005, Pratt2010},

\begin{equation}
	K/K_{500}= 1.47 \left(R/r_{500}\right)^{1.1} \label{eq.K}
\end{equation}
Here, $K_{500} = 106~ {\rm keV cm^{2}}(M_{500}/10^{14}M_\odot)^{2/3}(1/{f_{b}})^{2/3} E(z)^{-2/3}$,
where $f_b$ is the baryon fraction and $E(z)$ 
is the ratio of the Hubble constant at redshift $z$ with its present value.
Using the two-dimensional temperature slope and the three-dimensional density slope, we obtain the entropy slope of 
  0.81$\pm$0.25  for the relaxed arms  (table \ref{tab:powerlaw2}).
These values do not contradict the theoretical slope of 1.1.  
We also calculated $K$ using the weighted averages of $kT_{\rm ICM}$ shown in table \ref{tab:kt} and the power-law density profiles and got similar entropy profiles.
 As shown in figure \ref{fig:K},
from $\sim r_{500}$ to $r_{200}$,  the derived entropy profiles for the relaxed and W arms are close to the theoretical expectation (equation \ref{eq.K}). Here, we assume $f_{b}=0.15$, based on the results described in section \ref{sec:fb}.
Thus, the ICM in the Perseus cluster outskirts is heated by the gravity as expected from numerical simulations.

\subsection{Possible accretions from the filaments of the large-scale structure}

\begin{figure}
\centerline{		\includegraphics[width=8cm]{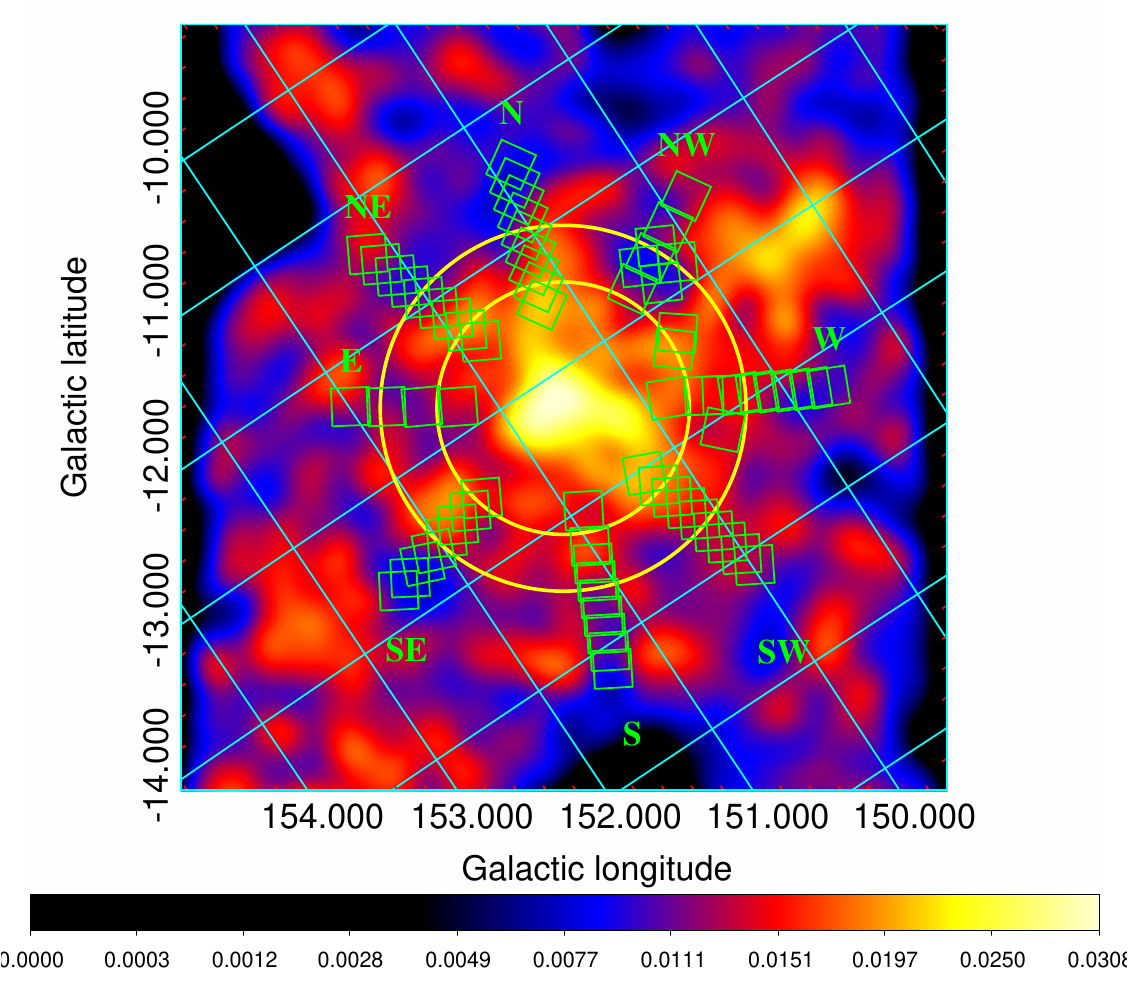}
}
\caption{Distribution of galaxies around the Perseus cluster, from SDSS photometric data within the redshift range of 3000 km/s,
with a smoothing scale (full-width half maximum) of 20$^\prime$.  The meanings of the squares and circles are the same as in figure \ref{fig:clusterimages}.
   {Alt text:A 2D smoothed density map of galaxies in Galactic coordinates centered on the Perseus cluster, with color shading indicating surface density. } }
\label{fig:galaxies}
\end{figure}

With the three-dimensional numerical simulations, \citet{simvirshock} found shocks, "virial shocks", located near the virial radius with Mach numbers of 2.5--4 toward filaments in relaxed clusters.
As accretion occurs along a filament, the infalling gas forms a shock inside the virial radius.
Inside the shock, the ICM density toward filaments flattens and is brighter than in the other directions.

The surface brightness map of galaxies with the K-band \citep{galaxies} and the distribution of galaxy clusters \citep{PPSC} indicate that the two filaments extend northwest and west from the Perseus cluster.
Figure \ref{fig:galaxies} shows the density distribution of galaxies with SDSS. 
Here, galaxies are selected using photometric redshifts spanning in
the range of $3000\rm{km ~s^{-1}}$ \citep{Ichikawa13}.
The galaxy density is higher in the W to NW directions than in the other directions.
This high galaxy density region likely connects to the filaments.

\citet{Urban14} and \citet{Zhu2021} found the discontinuities in the temperature and density jumps at $\sim 82^\prime$ along the NW arm.
They found that the ICM temperature drops from 2 keV to 1 keV,  corresponding to the Mach number, $M=1.9$, and discussed that the jump is caused by a "virial shock".
\citet{Walker2022}  analyzed XMM observations in the W directions of the Perseus cluster. They identified two X-ray surface brightness edges at 1.2 and 1.7  Mpc (51$^\prime$ and 74$^\prime$).
They proposed that these two edges are cold fronts created by a binary cluster merger,
although they cannot rule out that the outer edge results from stray light.
With the Suzaku pointings along the W arm, they found a temperature jump from 4 keV at $\sim 70^\prime$ to $\sim 6$ keV beyond 75$^\prime$, which does not contradict our temperature measurements.
Our spectra are extracted over the FOV and are not sensitive to these edge-like structures.
Nevertheless, we detected a sudden drop in EM$_{\rm ICM}$ between 81$^\prime$ and 88$^\prime$ pointings along the NW arm, while those along the W arm are continuous out to 91$^\prime$.
As shown in figure \ref{fig:spec},  the ICM brightness is quite different between the spectra at $r=81^\prime$ to $r=88^\prime$ along the NW arm, and at $r=88^\prime$, the ICM contribution is almost hidden by the HG component.
Therefore, it is challenging to detect ICM and constrain $kT_{\rm ICM}$ beyond the jump and distinguish whether this is a cold front or a virial shock.

The higher and flatter EM$_{\rm ICM}$ along the W and NW arms is consistent with the prediction from
the numerical simulations by \citet{simvirshock}. \citet{Walker2022} analyzed the surface brightness fluctuation of the western part of the Perseus cluster from 20$^\prime$ to 80$^\prime$ using the XMM-Newton observations and concluded that the clumping factor 
reaches up to 1.08 beyond $r_{500}$. Therefore, the gas-clumping effect is relatively mild and may not lead to an overestimation of the density measurements along these arms.

The pointing at 76$^\prime$ or 1.7 Mpc (RA = 48.3$^\circ$, DEC = 41.3$^\circ$ or $l=149.6^\circ, b=-14.1^\circ$), located south of the main W arm (DEC=41.6$^\circ$), shows lower $kT_{\rm ICM}$ and higher EM$_{\rm ICM}$ than those of the W arms.
\citet{Sasaki2015} and \citet{Sasaki2016} found group-scale X-ray enhancements around the dark matter subhalos detected with weak-lensing observations of the Coma cluster
by \citet{Okabe14b}. Their temperatures are lower than those of the surrounding ICM.
The X-ray enhancement at 76$^\prime$ may be caused by recently accreted substructures from the western filament.

 \subsection{Bias in temperature, density, and metal abundance measurements in the cluster outskirts}

The HG component can bias the temperature and density measurements of the ICM in the cluster outskirts.
In particular, a bright component with $\sim$ 0.8 keV is also seen in the directions towards the eROSITA bubble \citep{Gupta2022,Halo22}.
As described in section \ref{sec:sys}, when we fit the spectra without the HG component and the excess O\,\emissiontype {VII} line, allowing $kT_{\rm MWH}$ to vary, even at $\sim r_{500}$, the $kT_{\rm ICM}$ sometimes underestimated by $\sim 1$ keV at most.
 Furthermore, the HG component shows some spatial variation. 
 If we underestimate the contribution of the HG component, the remaining Fe-L peak may cause an underestimation of $kT_{\rm ICM}$.
 
\citet{Paper I} reported that the MWH component obtained around the solar maximum probably contaminated the heliospheric SWCX emissions. We also detected excess  O\,\emissiontype {VII} line emissions from the Perseus cluster observations (see appendix \ref{soft} for details).
Without this line component in spectral fittings, we may underestimate the temperature and overestimate the emission measure of the MWH component.  Then, we may also underestimate $kT_{\rm HG}$, which gives some systematic uncertainties in $kT_{\rm ICM}$  and EM$_{\rm ICM}$.
Furthermore,
when analyzing the spectra of the ICM component in the cluster outskirts, using another observation to constrain the LHB and MWH components is inappropriate.
For example,  since ROSAT observations are performed around the solar maximum, some contamination may be caused by the solar activity \citep{Uprety2016}.

\subsection{Origin of the supervirial temperature component}

\begin{figure*}
\centerline{\includegraphics[width=16.cm]{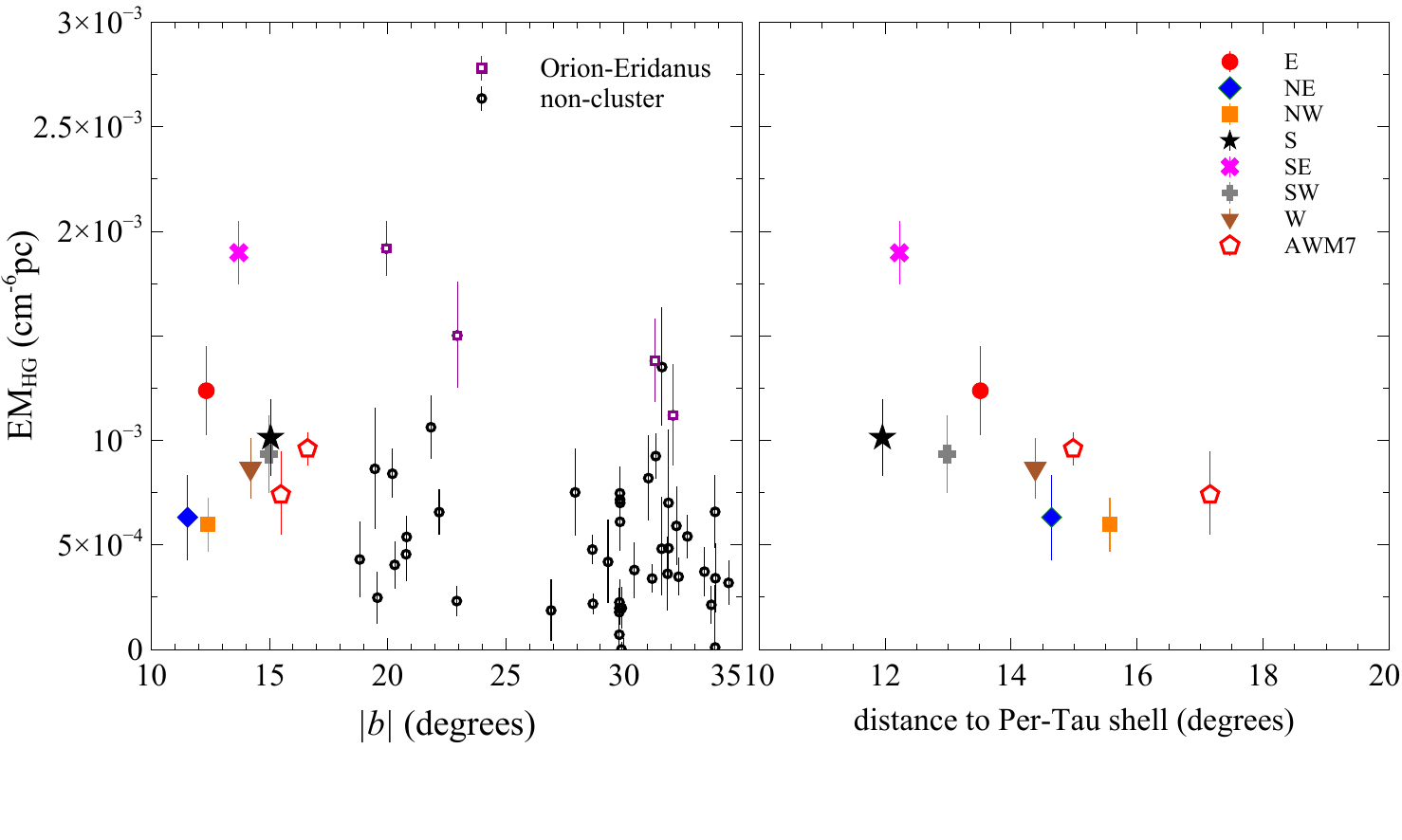}}
\caption{EM$_{\rm HG}$ of the outskirts of the Perseus (horizontal lines in Figure \ref{fig:08}) and AWM7 clusters plotted against the Galactic longitude (left panel) and the distance from the center of the Per-Tau shell (right panel).  Hexagons are the two regions at the outskirts of the AWM7 cluster.  In the left panel, those of the non-cluster sample (Model-08+OVII from  \citet{SugiyamaMWH2} ; the open circles; the open squares for the Orion-Eridanus superbubble) are also plotted.   {Alt text: Two-panel plot showing the emission measure of the hot galactic component: the left panel plots them against Galactic longitude, including data from clusters and non-cluster regions; the right panel shows the same values plotted against the distance to the Per-Tau shell. Hexagons indicate AWM7 outskirts, and symbols distinguish different observation fields.}  }
\label{fig:em08vslb}
\end{figure*}

\citet{SugiyamaMWH2} analyzed the 130 Suzaku observations outside the eROSITA bubble and reported that the HG component tends to be brighter towards the lower Galactic latitude, although there is a significant scatter at a given latitude. Based on the scatter in their emission measure, they discussed a possible relation with the stellar feedback.
We will refer to this sample as the non-cluster sample.
The left panel of figure \ref{fig:em08vslb} shows EM$_{\rm HG}$ (the horizontal lines in figure \ref{fig:08}) at the outskirts of the Perseus cluster and the AWM7 cluster, plotted against the absolute value of the Galactic latitude, $|b|$, with the non-cluster sample at $|b|<35^\circ$.
Except for the SE and E arms, the EM$_{\rm HG}$ matches those with the highest EM$_{\rm HG}$ of the non-cluster sample.
The EM$_{\rm HG}$ for the SE and E arms are about a factor of two higher and comparable to those of the brightest region of the Orion-Eridanus superbubble.
This bubble may be a nearby ($\sim$ 400 pc) large cavity extending from the Ori OB1 association
\citep{Reynolds79, Burrows1993, Snowden1995}.
The agreement of the EM$_{\rm HG}$ at the outskirts of the Perseus cluster with the non-cluster sample supports a similar origin for the HG component.

From the Perseus cluster outskirts observations, we can conclude that the HG component extends at least several degrees and varies slowly.
The spectral shape of the HG component resembles those of M-type stars. 
However, we expect a much smoother distribution if this component comes from stars. 
In the southwest of the Perseus cluster, there is the Perseus OB2 association, which has a typical age of 2--5 Myr \citep{Perstar}. Although the observed stellar dynamics do not support an expanding bubble, \citet{Pershell} studied the three-dimensional structure of the
molecular gas and dust and proposed a near-spherical shell extending over 30$^\circ$ in the sky between the Taurus and Perseus molecular clouds. They also found a diffuse X-ray enhancement near the center of the shell with ROSAT and eROSITA.
The Perseus and AWM7 clusters are close to the edge of this possible bubble. The right panel of Figure \ref{fig:em08vslb} shows EM$_{\rm HG}$ at the outskirts of the Perseus cluster, plotted against the distance to the center of the Perseus-Taurus shell. There may be a hint of enhancement towards the shell center. 
The enhancement of EM$_{\rm HG}$ towards the SE and E arms may be related to the possible bubble from the star formation in the Perseus star clusters.

 \section{Summary and conclusion}

 We analyzed 65 Suzaku observations beyond   1 Mpc ($\sim$0.8 $r_{500}$)   of the Perseus cluster in eight azimuthal directions.  The two directions are towards the filaments of the large-scale structure in the Universe.
 The modeling of the soft X-ray background is crucial to studying the thermodynamic properties in the ICM in the cluster outskirts.
Beyond 95$^\prime$, or 1.6 $r_{500}$,
there is a one keV CIE plasma component whose emission measure is relatively flat along each arm but shows some azimuthal dependence.
Considering the possible spatial variation of this component, we fitted the spectra extracted over the FOV of each observation within the virial radius.
We also need the additional O\,\emissiontype {VII}  He$\alpha$ line, possibly from heliospheric SWCX, to fit the spectra obtained around the solar maximum.

Excluding the directions towards the filaments, or relaxed arms, at a given radius, the scatter in $kT_{\rm ICM}$ and EM$_{\rm ICM}$ is relatively small.
The radial profile of $kT_{\rm ICM}$ for the  relaxed arms   is represented by a power-law with an index of  -0.67$\pm$0.25.  
The radial profile of EM$_{\rm ICM}$ at   $r=49^\prime$--$101^\prime$ ($\sim 0.8~ r_{500}$ to $\sim 1.7~r_{500}$)    is also  described with a power-law with an index of   -3.5$\pm$0.1.   This slope corresponds to a density slope of -2.21$\pm$0.06.  
Assuming hydrostatic equilibrium, the radial profiles of $kT_{\rm ICM}$ and EM$_{\rm ICM}$ give $r_{500}=59^\prime$ (1.3 Mpc)  and $M_{500}=6.4\times 10^{14} M_\odot$.
The integrated gas fraction, the ratio of the integrated ICM mass to the hydrostatic mass, is 0.13$\pm$0.01 and 0.18$\pm$0.02 at $r_{500}$ and $r_{200}$, respectively.   
Considering the stellar mass fraction, the baryon fraction agrees with the cosmic baryon fraction of \citet{Planck18}.
The pressure profile for the relaxed arms is close to the average profile by \citet{Planck2013}.
The entropy profile is also close to the theoretical expectation from numerical simulations assuming the gravitational heating.
These results indicate that the ICM at the cluster outskirts is quite regular and close to hydrostatic equilibrium.

Within a few times $r_{500}$, the distribution of galaxies elongates toward the northwest direction, possibly connecting to the two filaments beyond the NW and W arms.
In contrast to the relaxed arms, these two arms have higher EM$_{\rm ICM}$ by a factor of 1.5--2 than those for the relaxed arms at a given radius, as expected from the numerical simulations.

The temperature and surface brightness of the HG component beyond 1.6 $r_{500}$ are consistent with those found in regions without galaxy clusters observed with Suzaku. This component sometimes gives significant biases to the measurements of temperature and  density of the ICM in cluster outskirts. 
The spatial variation of the HG component suggests that this component is related to the stellar feedback in our Galaxy.

\appendix
\small
\chapter{Observational log}
\label{appendix:observation_info}
Table \ref{tab:obslog} lists the Suzaku observations used in this work.

\begin{longtable}{llrrrccr}
    \caption{Observational log of the cluster sample}
    \label{tab:obslog} \\
    \hline
       Target Name$^*$  & label{$^\dagger$} &  RA$^\ddagger$  & DEC$^\|$ & offset$^\S$ & Sequence{\(^\#\)} & date$^{**}$   & Time$^\dagger\dagger$\\ \hline
    \endfirsthead
     Target Name$^*$  & label{$^\dagger$} &  RA$^\ddagger$  & DEC$^\|$ & offset$^\S$ & Sequence{\(^\#\)} & date$^{**}$   & Time$^\dagger\dagger$ \\ \hline
        \endhead
    \hline
    \endfoot
    \hline
    \multicolumn{6}{l}{\footnotemark[$*$]Name given to the pointed target in the archive} \\
      \multicolumn{6}{l}{\footnotemark[$\dagger$]Azimuthal direction used in Figure \ref{fig:clusterimages}} \\
   \multicolumn{6}{l}{\footnotemark[$\ddagger$]	The Right Ascension (J2000) of the pointing position in units of degrees} \\
   \multicolumn{6}{l}{\footnotemark[$\|$]	The Declination (J2000) of the pointing position in units of degrees} \\
      \multicolumn{6}{l}{\footnotemark[$\S$]	Offset angle from the cluster center in units of arcmin} \\
    \multicolumn{6}{l}{\footnotemark[$\#$]	Sequence numbers of the Suzaku archive} \\
    
      \multicolumn{6}{l}{\footnotemark[$**$]	Observation start date} \\
       \multicolumn{6}{l}{\footnotemark[$\dagger\dagger$]	Exposure time in units of ks} \\
    %
    \endlastfoot
 ABELL 426 E3 & E &  51.065 & 41.5110& 50.2& 804058010 & 2009-07-29& 10.9 \\ 
 ABELL 426 E4 & E &  51.439 & 41.5068& 67.0& 804059010 & 2009-07-30& 21.5 \\ 
 ABELL 426 E5 & E &  51.812 & 41.5020& 83.8& 804060010 & 2009-07-30& 21.2 \\ 
 ABELL 426 E6 & E &  52.185 & 41.4955& 100.6& 804061010 & 2009-07-31& 31.5 \\ 
 ABELL 426 NE3 & NE &  50.831 & 42.0254& 50.1& 806126010 & 2011-09-05& 10.3 \\ 
 ABELL 426 NE3.5 & NE &  50.984 & 42.1082& 58.6& 806127010 & 2011-09-05& 10.2 \\ 
 ABELL 426 NE4 & NE &  51.133 & 42.1979& 67.1& 806128010 & 2011-09-04& 10.2 \\ 
 ABELL 426 NE4.5 & NE &  51.285 & 42.2822& 75.5& 806129010 & 2011-09-05& 10.3 \\ 
 ABELL 426 NE5 & NE &  51.436 & 42.3617& 83.7& 806130010 & 2011-09-06& 16.7 \\ 
 ABELL 426 NE5.5 & NE &  51.590 & 42.4448& 92.2& 806131010 & 2011-09-06& 14.2 \\ 
 ABELL 426 NE6 & NE &  51.740 & 42.5297& 100.6& 806132010 & 2011-09-06& 11.1 \\ 
 ABELL 426 NE6.5 & NE &  51.897 & 42.6152& 109.2& 806133010 & 2011-09-07& 9.7 \\ 
 ABELL 426 NE7 & NE &  52.047 & 42.6926& 117.3& 806134010 & 2011-09-07& 10.8 \\ 
 ABELL 426 NNE3 & N &  50.189 & 42.3197& 49.6& 806138010 & 2011-08-31& 9.4 \\ 
 ABELL 426 NNE3.5 & N &  50.228 & 42.4583& 58.1& 806139010 & 2011-08-31& 10.0 \\ 
 ABELL 426 NNE4 & N &  50.277 & 42.5909& 66.3& 806140010 & 2011-09-01& 11.7 \\ 
 ABELL 426 NNE4.5 & N &  50.314 & 42.7296& 74.8& 806141010 & 2011-09-02& 10.9 \\ 
 ABELL 426 NNE5 & N &  50.359 & 42.8663& 83.2& 806142010 & 2011-09-02& 16.1 \\ 
  ABELL 426 NNE5.5 & N &  50.399 & 43.0029& 91.6& 806143010 & 2011-09-02& 16.2 \\ 
 ABELL 426 NNE6.5 & N &  50.487 & 43.2740& 108.3& 806145010 & 2011-09-03& 13.3 \\ 
 ABELL 426 NNE7 & N &  50.528 & 43.4162& 117.0& 806146010 & 2011-09-03& 10.9 \\ 
 ABELL 426 NNE6 & N &  50.442 & 43.1400& 100.1& 806144010 & 2011-09-03& 10.3 \\ 
 ABELL 426 N4 & NW &  49.229 & 42.4427& 64.3& 804066010 & 2009-08-19& 25.1 \\ 
 A426VIR S & NW &  48.890 & 42.5005& 75.7& 808085010 & 2014-02-20& 25.5 \\ 
 A426VIR E & NW &  49.111 & 42.6307& 76.7& 808086010 & 2014-02-21& 25.1 \\ 
 ABELL 426 N5 & NW &  49.037 & 42.6835& 81.1& 804067010 & 2009-08-20& 23.3 \\ 
 A426VIR W & NW &  48.718 & 42.6588& 87.9& 808088010 & 2014-02-23& 24.5 \\ 
 A426VIR N & NW &  48.939 & 42.7906& 88.8& 808087010 & 2014-02-22& 21.7 \\ 
 ABELL 426 N6 & NW &  48.843 & 42.9236& 97.8& 804068010 & 2009-08-20& 35.7 \\ 
 ABELL 426 N7 & NW &  48.648 & 43.1648& 114.6& 804069010 & 2009-08-21& 34.9 \\ 
 ABELL 426 S3 & S &  49.749 & 40.7028& 49.4& 805098010 & 2010-08-17& 11.0 \\ 
 ABELL 426 S4 & S &  49.683 & 40.4247& 66.4& 805099010 & 2010-08-17& 12.2 \\ 
 ABELL 426 S4.5 & S &  49.650 & 40.2869& 74.8& 805110010 & 2010-08-18& 8.9 \\ 
 ABELL 426 S5 & S &  49.619 & 40.1482& 83.2& 805100010 & 2010-08-18& 10.5 \\ 
 ABELL 426 S5.5 & S &  49.586 & 40.0129& 91.5& 805111010 & 2010-08-18& 11.0 \\ 
 ABELL 426 S6 & S &  49.554 & 39.8750& 99.9& 805101010 & 2010-08-18& 16.2 \\ 
 ABELL 426 S6.5 & S &  49.522 & 39.7382& 108.2& 805112010 & 2010-08-19& 15.1 \\ 
 ABELL 426 S7 & S &  49.490 & 39.6000& 116.7& 805102010 & 2010-08-19& 16.9 \\ 
 ABELL 426 S7.5 & S &  49.459 & 39.4611& 125.1& 805113010 & 2010-08-20& 14.6 \\ 
 ABELL 426 SE3.5 & SE &  50.804 & 40.7976& 57.8& 806115010 & 2011-08-25& 11.5 \\ 
 ABELL 426 SE4 & SE &  50.920 & 40.6946& 66.0& 806116010 & 2011-08-25& 10.7 \\ 
 ABELL 426 SE4.5 & SE &  51.047 & 40.5864& 74.6& 806117010 & 2011-08-25& 11.9 \\ 
 ABELL 426 SE5 & SE &  51.163 & 40.4865& 82.6& 806118010 & 2011-08-26& 16.3 \\ 
 ABELL 426 SE5.5 & SE &  51.289 & 40.3763& 91.4& 806119010 & 2011-08-26& 16.2 \\ 
 ABELL 426 SE6 & SE &  51.404 & 40.2751& 99.4& 806120010 & 2011-08-26& 9.7 \\ 
 ABELL 426 SE6.5 & SE &  51.531 & 40.1661& 108.2& 806121010 & 2011-08-27& 9.5 \\ 
 ABELL 426 SE7 & SE &  51.645 & 40.0628& 116.3& 806122010 & 2011-08-27& 10.2 \\ 
 ABELL 426 SW3 & SW &  49.115 & 40.9879& 49.0& 806102010 & 2011-08-28& 11.8 \\ 
 ABELL 426 SW3.5 & SW &  48.969 & 40.8952& 57.6& 806103010 & 2011-08-28& 10.2 \\ 
 ABELL 426 SW4 & SW &  48.834 & 40.8048& 65.8& 806104010 & 2011-08-28& 13.1 \\ 
 ABELL 426 SW4.5 & SW &  48.687 & 40.7137& 74.4& 806105010 & 2011-08-28& 10.6 \\ 
 ABELL 426 SW5 & SW &  48.554 & 40.6247& 82.5& 806106010 & 2011-08-29& 16.5 \\ 
 ABELL 426 SW5.5 & SW &  48.408 & 40.5297& 91.3& 806107010 & 2011-08-29& 15.0 \\ 
 ABELL 426 SW6 & SW &  48.277 & 40.4394& 99.4& 806108010 & 2011-08-29& 11.4 \\ 
 ABELL 426 SW6.5 & SW &  48.131 & 40.3470& 108.0& 806109010 & 2011-08-30& 11.3 \\ 
 ABELL 426 SW7 & SW &  47.998 & 40.2589& 116.1& 806110010 & 2011-08-30& 10.3 \\ 
 ABELL 426 W3 & W &  48.859 & 41.5755& 49.0& 805105010 & 2010-08-20& 10.9 \\ 
 MRK 1073 & W &  48.759 & 41.9794& 60.1& 701007020 & 2007-02-04& 38.4 \\ 
 MRK 1073 & W &  48.756 & 42.0893& 63.5& 701007010 & 2006-08-02& 2.9 \\ 
 ABELL 426 W4 & W &  48.484 & 41.5928& 65.8& 805106010 & 2010-08-21& 10.1 \\ 
 ABELL 426 W4.5 & W &  48.299 & 41.5966& 74.1& 805114010 & 2010-08-21& 10.7 \\ 
 B3 0309+411 & W &  48.264 & 41.3383& 76.4& 706036010 & 2012-02-19& 101.7 \\ 
 ABELL 426 W5 & W &  48.113 & 41.6030& 82.5& 805107010 & 2010-08-21& 9.2 \\ 
 ABELL 426 W5.5 & W &  47.926 & 41.6103& 90.9& 805115010 & 2010-08-21& 9.7 \\ 
 ABELL 426 W6 & W &  47.738 & 41.6154& 99.3& 805108010 & 2010-08-22& 15.0 \\ 
 ABELL 426 W6.5 & W &  47.552 & 41.6199& 107.7& 805116010 & 2010-08-22& 15.9 \\ 
 ABELL 426 W7 & W &  47.364 & 41.6278& 116.1& 805109010 & 2010-08-23& 15.2 \\ 
 ABELL 426 W7.5 & W &  47.178 & 41.6348& 124.5& 805117010 & 2010-08-23& 14.6 \\ 
 AWM7 BGD & NW &  41.677 & 42.4943& 102.9& 808027010 & 2014-02-19& 20.1 \\
  AWM7 SOUTH OFFSET & S &  43.203 & 40.6382& 59.1& 802045020 & 2008-02-23& 88.4 \\
 AWM7 SOUTH OFFSET & S &  43.203 & 40.6378& 59.2& 802045010 & 2008-01-29& 30.1 \\ 

\end{longtable}

\chapter{The NXB spectral fitting}
\label{sec:nxb}

\begin{figure}
    \centerline{\includegraphics[width=8cm]{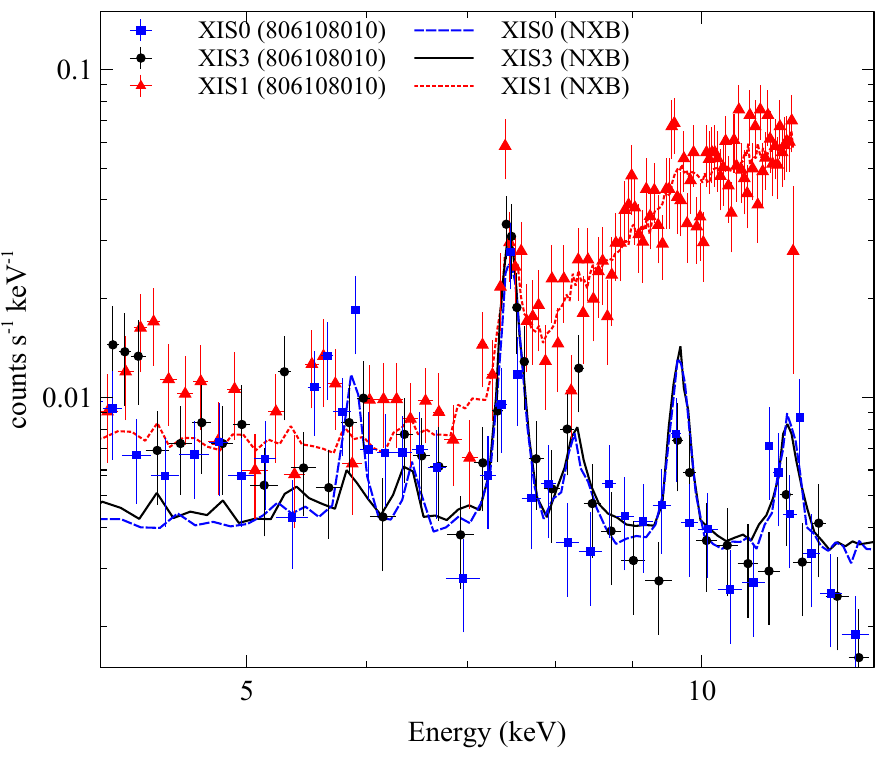}}
\caption{  The spectra of XIS0 (filled squares), XIS3 (filled circles), and XIS1 (filled triangles) from obsid= 806108010, with the corresponding NXB spectra (XIS0:dashed line, XIS3: solid line, XIS1: dashed line) {Alt text: Overlaid spectral curves with three sets of data points distinguished by marker shape. The vertical axis shows count rate per kiloelectronvolt, and the horizontal axis shows energy in kiloelectronvolt.}}
\label{fig:nxbspec0}
\end{figure}

The XIS spectral fittings are usually performed by subtracting the NXB spectra generated from the NXB database \citep{Tawa}, which is constructed by calculating a weighted sum of night-Earth observations over several months  using the revised cut-off rigidity (COR2). \citet{Tawa} estimated that the reproducibility of the NXB level in the 5--12 keV band is on the order of a few percent for each 5 ks exposure bin. However, their study was based on early Suzaku observations and may not be applicable to later data, especially those taken between 2010 and 2014 during the solar maximum \citep{Yamamoto2020}.
  In analyzing the Suzaku data of the Perseus cluster, \citet{Zhu2021} modeled the NXB rather than subtracting it. %
  
We created NXB spectra for each observation from the Night-Earth database using {\it xisnxbgen},  accumulating over a period from 150 days before the start of each observation to 150 days after its end
  \citep{Tawa}.
Figure~\ref{fig:nxbspec0} compares the XIS spectra for OBSID=806108010 (SW 99$^\prime$) with the corresponding NXB database spectra.
 Above  7 keV, the NXB dominates the spectra.  The NXB levels from the sky observations and from the NXB database for the XIS0 and XIS3 detectors exhibit significant discrepancies at $>$ 10 keV.
This mismatch in the NXB normalization indicates that subtracting the NXB spectra, or performing a simultaneous fit of the sky spectra and the NXB database spectra, may not always be appropriate. 

\begin{figure}
    \centerline{\includegraphics[width=8cm]{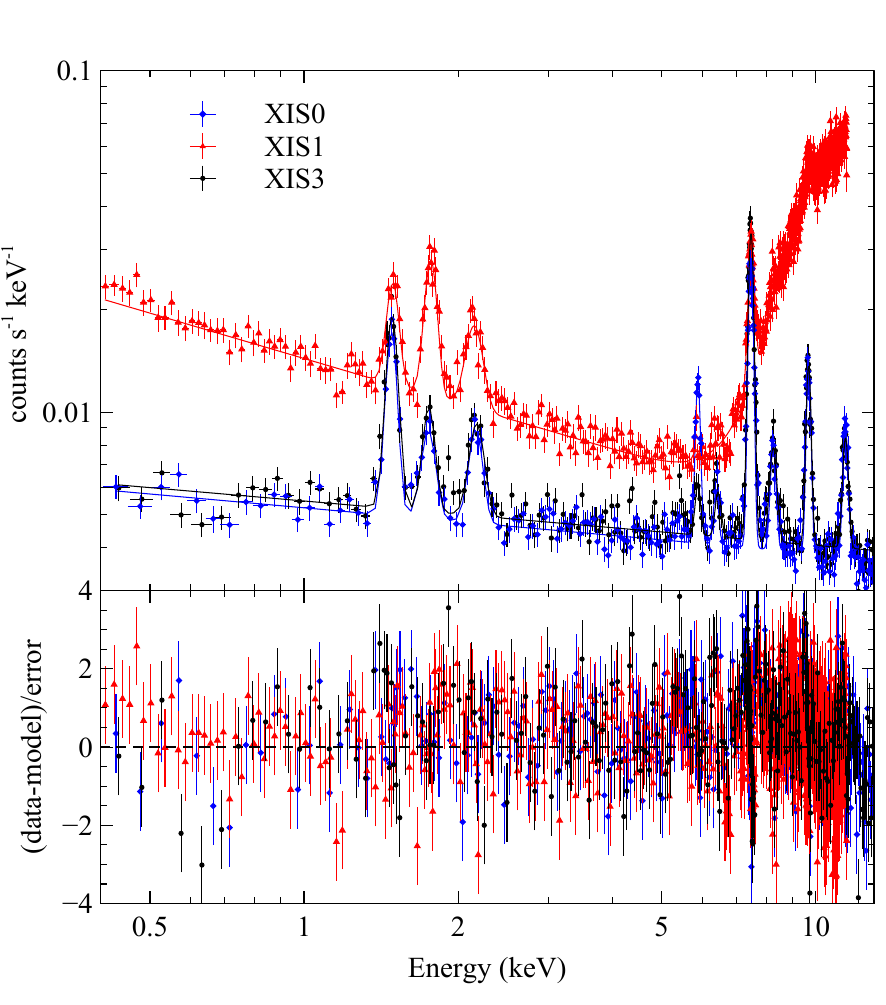}}
\caption{  The NXB spectra of XIS0 (blue diamonds), XIS1 (red triangles), and XIS3 (black circles) for obsid=806109010 (SW 108$^\prime$), fitted with a power-law and Gaussians lines. The upper panel displays the spectra and best-fit models, while the lower panel shows the residuals, defined as (data - model) divided by the statistical error. {Alt text:The upper panel shows the three overlaid spectra plotted with different colors and marks and a logarithmic vertical axis, with horizontal axis covering 0.4--13 keV.}} 
\label{fig:nxbspec}
\end{figure}

\begin{figure}
\centerline{   \includegraphics[width=4cm]{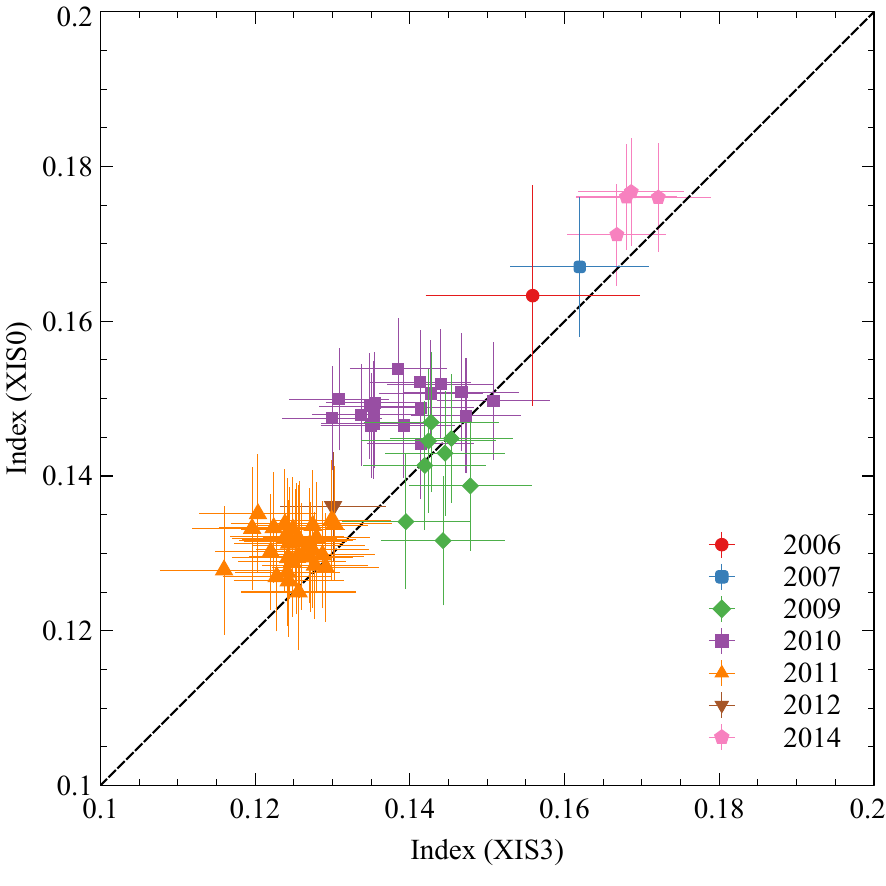}
     \includegraphics[width=4cm]{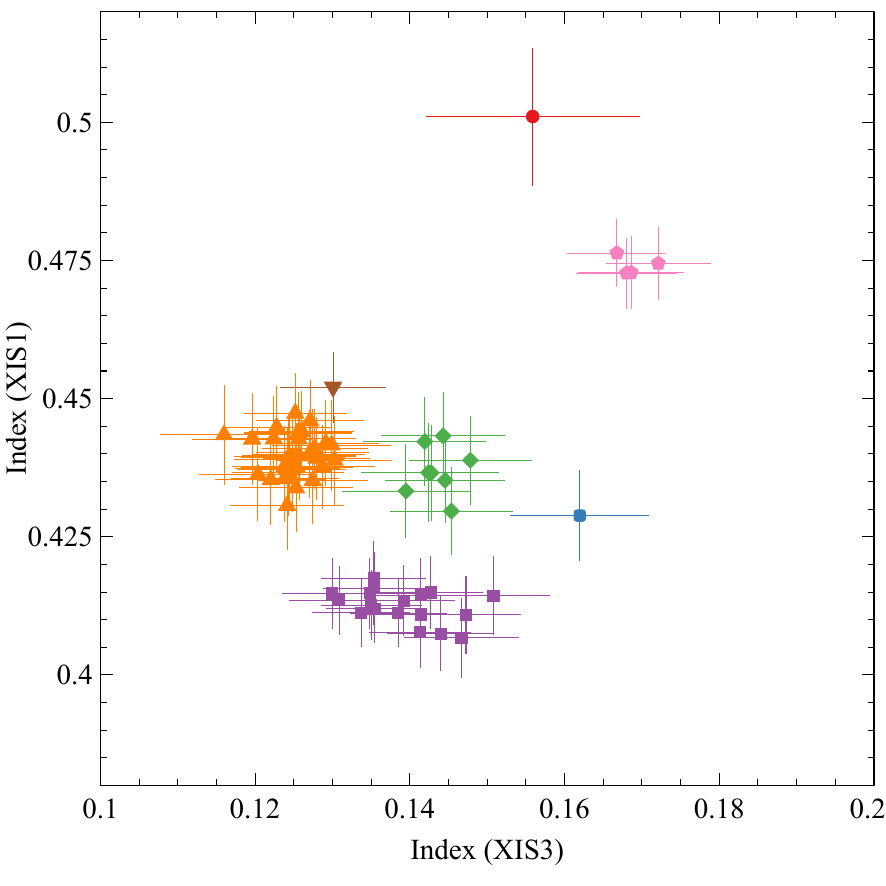}}
\caption{  The power-law indices for the NXB spectra of the XIS0 (left panel) and XIS1 (right panel) detectors plotted
against those of the XIS3.  Different symbols indicate different observation years. The dashed line in the left panel marks the equality between horizontal and vertical axis values.
{Alt text: Two scatter plots comparing the power-law indices of XIS0 (left) and XIS1 (right) with those of XIS3. Each point represents one observation, and marker styles distinguish observation years from 2006 to 2014.} }
\label{fig:nxb}
\end{figure}

Using a diagonal RMF file, we modeled the NXB spectra with a power-law and nine Gaussian lines, representing Al-K
$\alpha$,  Si-K$\alpha$,  Au-M$\alpha$,  Mn-K$\alpha$,  Mn-K$\beta$,  Ni-K$\alpha$,  Ni-K$\beta$
 Au-L$\alpha$,  and Au-L$\beta$
 \citep{Tawa}. To account for the broad excess above 5 keV in the XIS1 spectrum, we added a broad Gaussian centered at 11.4 keV with 
$\sigma$ =1.9 keV.  Each NXB spectrum was grouped to have a minimum of 30 counts per bin. 
We fitted the NXB spectra of the three detectors using 
$\chi^2$ statistic since the NXB spectra have Gaussian statistics. The normalizations of all components, the power-law index, and the centroid energies and widths of the Gaussians were allowed to vary. This model reproduces the observed NXB spectra reasonably well, as shown in Figure~\ref{fig:nxbspec}.
As presented in Figure~\ref{fig:nxb}, the power-law indices for XIS0 and XIS3 are  0.12--0.18 and show a strong correlation. Values from the same year are similar, likely due to overlapping data in the NXB database. In contrast, the indices for XIS1 are 0.4--0.5 and do not correlate well with those for XIS3. Possibly due to systematic uncertainties in the gain, the centroid energies of the Gaussians are slightly shifted from the theoretical values.

\begin{figure}
\centerline{ 
     \includegraphics[width=4cm]{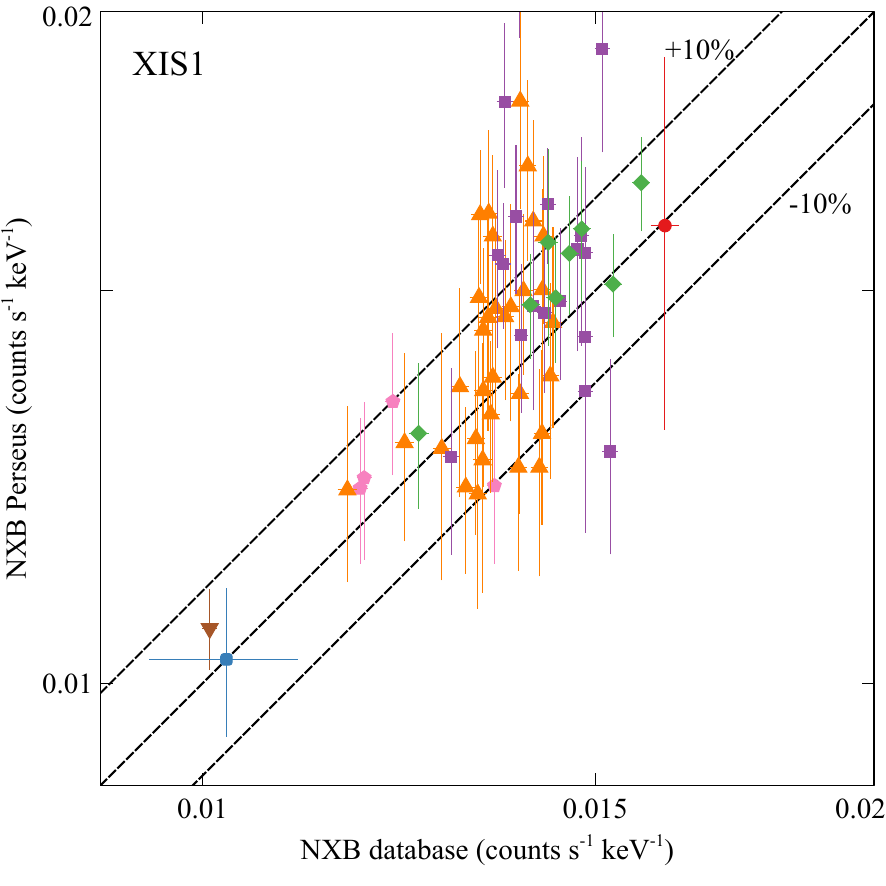}
        \includegraphics[width=4cm]{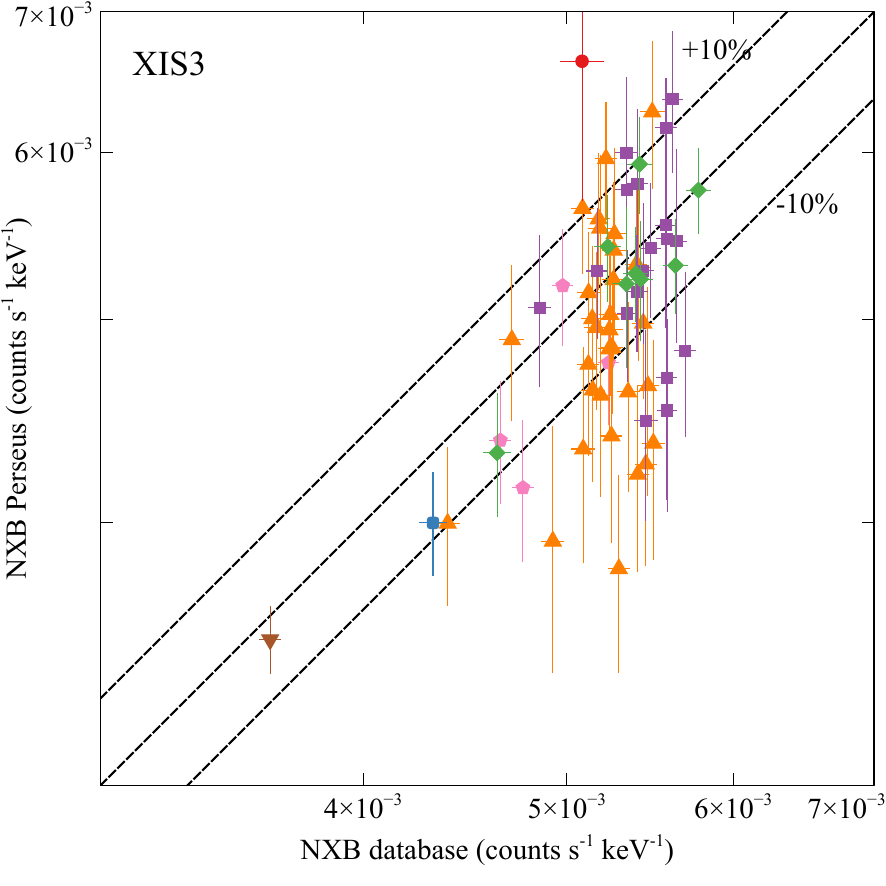}}
\caption{  The normalizations of the power-law component of NXB at 1 keV of the Perseus spectra of XIS1 (left) and XIS3 (right) fitted with the baseline model, plotted against those obtained from the corresponding NXB database spectra.  The three dashed lines represent -10\%, 0\%, and 10\% deviations of the vertical axis values from the horizontal axis values.
The meanings of the symbols are the same as in figure \ref{fig:nxb}. {Alt text: Two scatter plots comparing the 1 keV normalization of the NXB power-law component from the Perseus spectra (vertical axis) with that from the NXB database (horizontal axis), for XIS3 (left) and XIS1 (right).}}
\label{fig:nxbnorm}
\end{figure}

When fitting the sky spectra,  we incorporated the NXB model (a power law and Gaussians)  using the diagonal RMF file. The power-law indices, the centroid energies, and the Gaussian widths were fixed at the best-fit values determined from the corresponding NXB fits, while the normalizations of the power-law and Gaussians were allowed to vary. This model reproduces the observed Perseus spectra reasonably well, as shown in Figures~\ref{fig:bgd} and \ref{fig:spec}.
Figure~\ref{fig:nxbnorm} compares the normalization of the power-law component of the NXB at 1 keV derived from the Perseus data fitted with the baseline model (section \ref{sec:fits})  with that from the corresponding NXB spectra. In some cases, the two normalizations differ by a few tens of percent.  The 2009 data, taken at the solar minimum, show better consistency than observations taken after 2010, when solar activity increased.
We adopt these fits in the analysis presented in Section~\ref{sec:ana}.

\begin{figure}
   \centerline{\includegraphics[width=4cm]{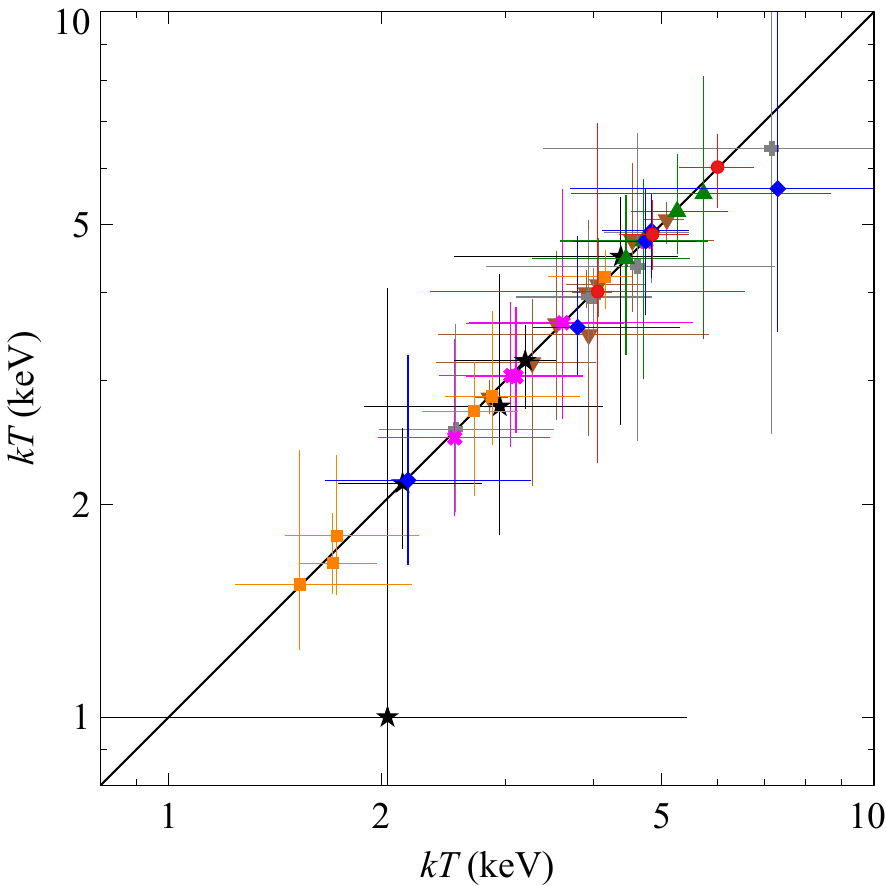}\includegraphics[width=4cm]{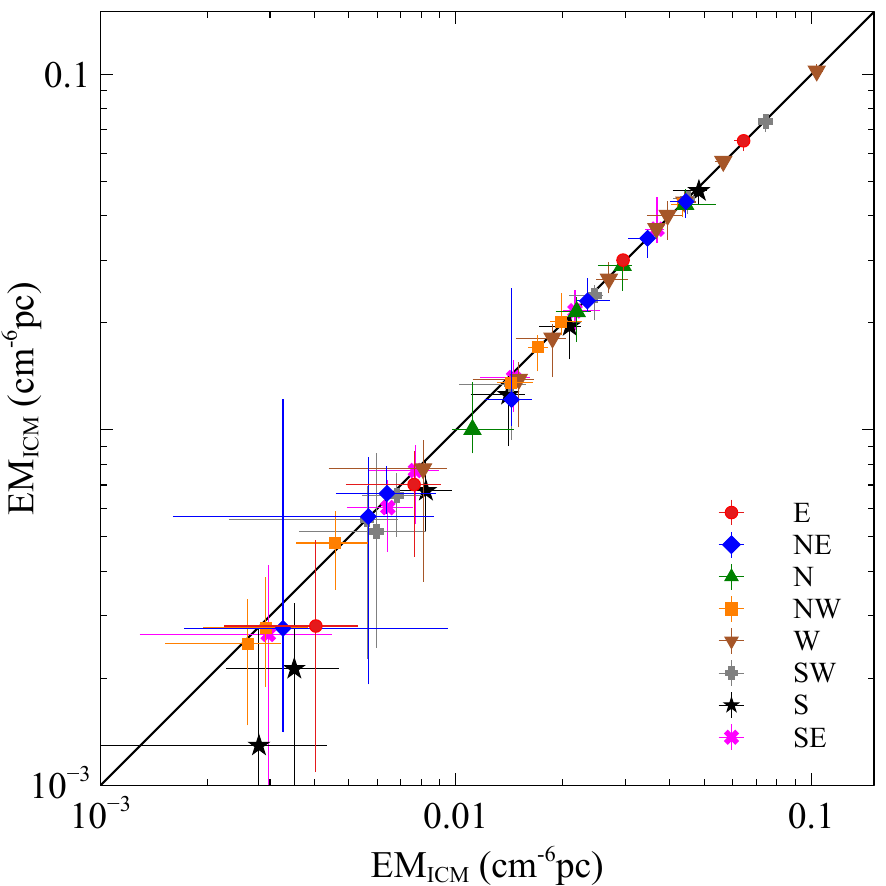}}
\caption{  (left panel)
Comparison of the ICM temperature, $kT_{\rm ICM}$ within 95$'$, derived with the power-law indices of the non-X-ray background (NXB)
 left free within the range allowed by the NXB spectra obtained from the NXB database, and those obtained in Section~\ref{sec:fits}, 
 where the indices were fixed at the best-fit values derived from the corresponding NXB spectra.
The meanings of the symbols and colors are the same as in Figure~\ref{fig:radial}.
The solid line indicates equality between the two values.
(right panel) Same as the left panel, but for the ICM emission measure, ${\rm EM}_{\rm ICM}$ within 105$'$.
{Alt text:  Two scatter plots comparing results from modified NXB models with the baseline fits for the Perseus cluster.  }} 
\label{fig:sysnxb}
\end{figure}

When the indices of the NXB power-law component are allowed to vary within the ranges (0.12--0.18 for XIS0 and XIS3 and 0.4--0.5 for XIS1),  the results remain consistent with those obtained using the fixed indices  as shown in figure \ref{fig:sysnxb}. The systematic uncertainties in the derived ICM temperature and emission measure associated with the NXB power-law index are much smaller than the other systematic uncertainties
discussed in section \ref{sec:sys}. 


\chapter{The normalization of the CXB component} \label{back}

\begin{figure}
    \centerline{\includegraphics[width=8cm]{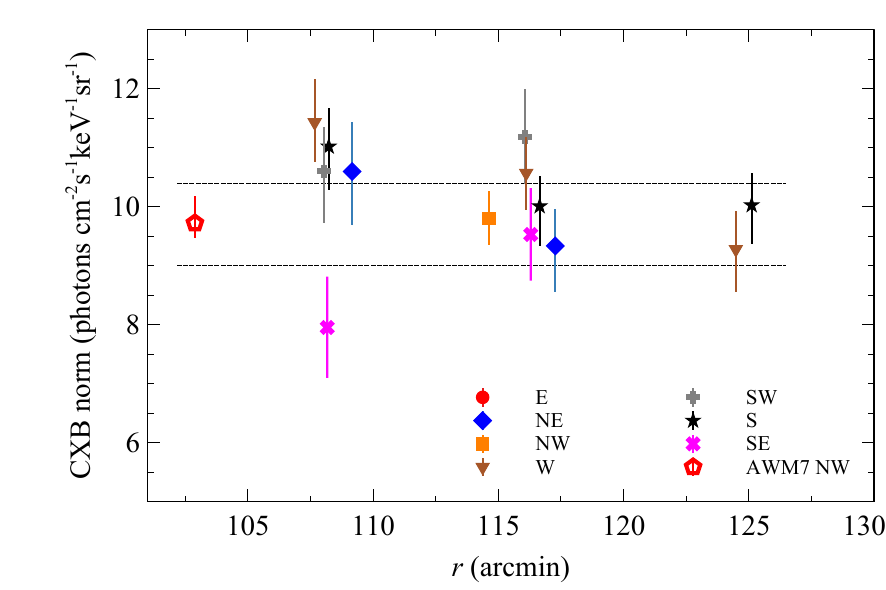}}
\caption{ The normalization of the power-law component at 1 keV for the CXB plotted against the distance from the cluster center. The horizontal lines show the adopted range of the CXB level. The meaning of marks and colors is the same as in figure \ref{fig:08}. {Alt text: A plot showing the CXB normalization at 1 keV as a function of projected radius for several pointing directions around the Perseus cluster and the background region of  AWM7. Horizontal dashed lines mark the adopted CXB range. } }
\label{fig:cxb}
\end{figure}

The normalization of the power-law component to model CXB depends on the flux level of the excluded point sources (e.g. \cite{Moretti2003}).
The CXB normalization of the stacked spectra of the 130 observations with Suzaku is $\sim$ 8.5 photons $\rm{cm^{-2}s^{-1}keV^{-1}sr^{-1}}$ at 1 keV \citep{Paper I}.
This value agrees with the threshold for excluding the point sources of $5.0\times 10^{-14}\rm{erg s^{-1}cm^{-2}}$ in the 2.0--10.0 energy range by \citet{Moretti2003}.
 Figure \ref{fig:cxb} shows that CXB normalizations with the background model at $r>102^\prime$ of the Perseus cluster and the AWM7 cluster are nearly constant.  
Their  weighted average  beyond 114$^\prime$ of the Perseus cluster is  9.7~  photons $\rm{cm^{-2}s^{-1}keV^{-1}sr^{-1}}$ at 1 keV.
 
If we extrapolate the results of \citet{Moretti2003}, this value corresponds to
the point source threshold of  $10^{-13}\rm{erg s^{-1}cm^{-2}}$.   
This difference is explained by the different exposure times of the two samples: $\sim$ 60 ks on average for the non-cluster sample, while the typical exposure time for the Perseus cluster is 10--20 ks.

The fluctuation of the CXB normalization also depends on the threshold flux due
to unresolved point sources.
This fluctuation is proportional to $\Omega_{0.01}^{-1/2}$, where
$\Omega_{0.01}$ is the solid angle of the corresponding region in units of 10$^{-2}~\rm{deg}^2$.
Based on \citet{Moretti2003}, \citet{Bautz09} estimated the RMS fluctuations 
at 2--10 keV is $\sim$ 20\% $\times   \Omega_{0.01}^{-1/2}$ for the threshold flux of $1.0\times 10^{-13}\rm{erg s^{-1}cm^{-2}}$.
Then, for the Suzaku FOV  (0.09 deg$^2$) , the estimated fluctuation is
 7 ~\% for this threshold flux. 
Therefore, in this paper, we allowed the normalization of the CXB to vary within a limited range of 
 9.0-10.4  ~photons $\rm{cm^{-2}s^{-1}keV^{-1}sr^{-1}}$ at 1 keV.

\chapter{The soft X-ray background}
\label{soft}

\begin{figure}
    \centerline{\includegraphics[width=8cm]{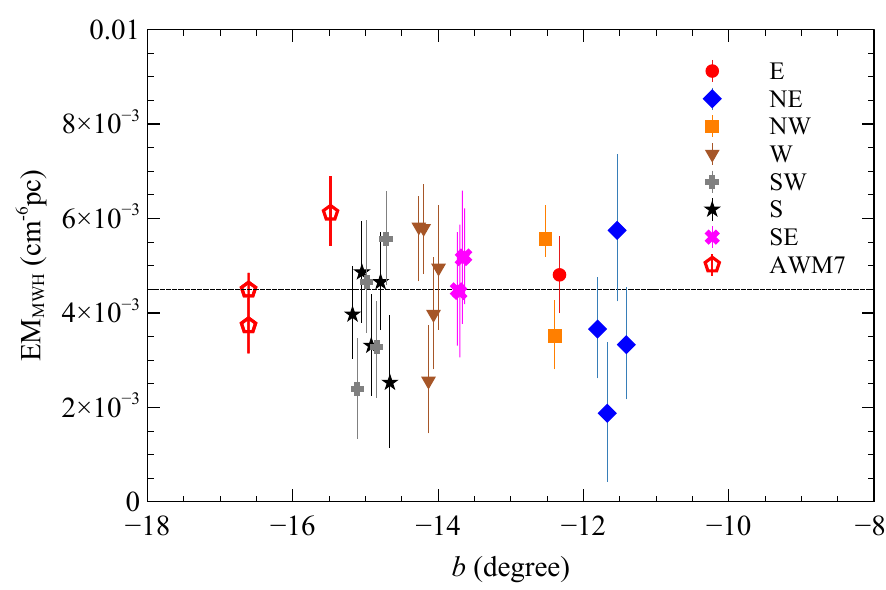}}
\caption{EM$_{\rm MWH}$ plotted against the Galactic latitude, $b$. The horizontal line shows the weighted average. The meaning of marks and colors is the same as in figure \ref{fig:08}   {Alt text:A plot showing the emission measure of the Milky Way Halo as a function of the Galactic latitude. Each mark corresponds to a different pointing direction.}  }
\label{fig:mwh}
\end{figure}

\citet{Paper I} reported that the spatial distribution of the emission measure of the MWH (hereafter EM$_{\rm MWH}$) is well represented by a sum of disk-like and spherical morphology components.
  The scatter in EM$_{\rm MWH}$ is small towards the anti-Galactic center.
  Figure \ref{fig:mwh} shows EM$_{\rm MWH}$ values in the background regions plotted against the Galactic latitude, $b$.
EM$_{\rm MWH}$ is nearly constant, with a weighted average of 4. 5 $\times 10^{-3}\rm{cm^{-6}pc}$. This value agrees with the EM$_{\rm MWH}$ for the non-cluster sample at similar Galactic longitude ($l$) and latitude ($b$) and is consistent with the best-fit model for the disk-like and spherical morphology model of \citet{Paper I}.

\begin{figure*}
   \centerline{\includegraphics[width=16cm]{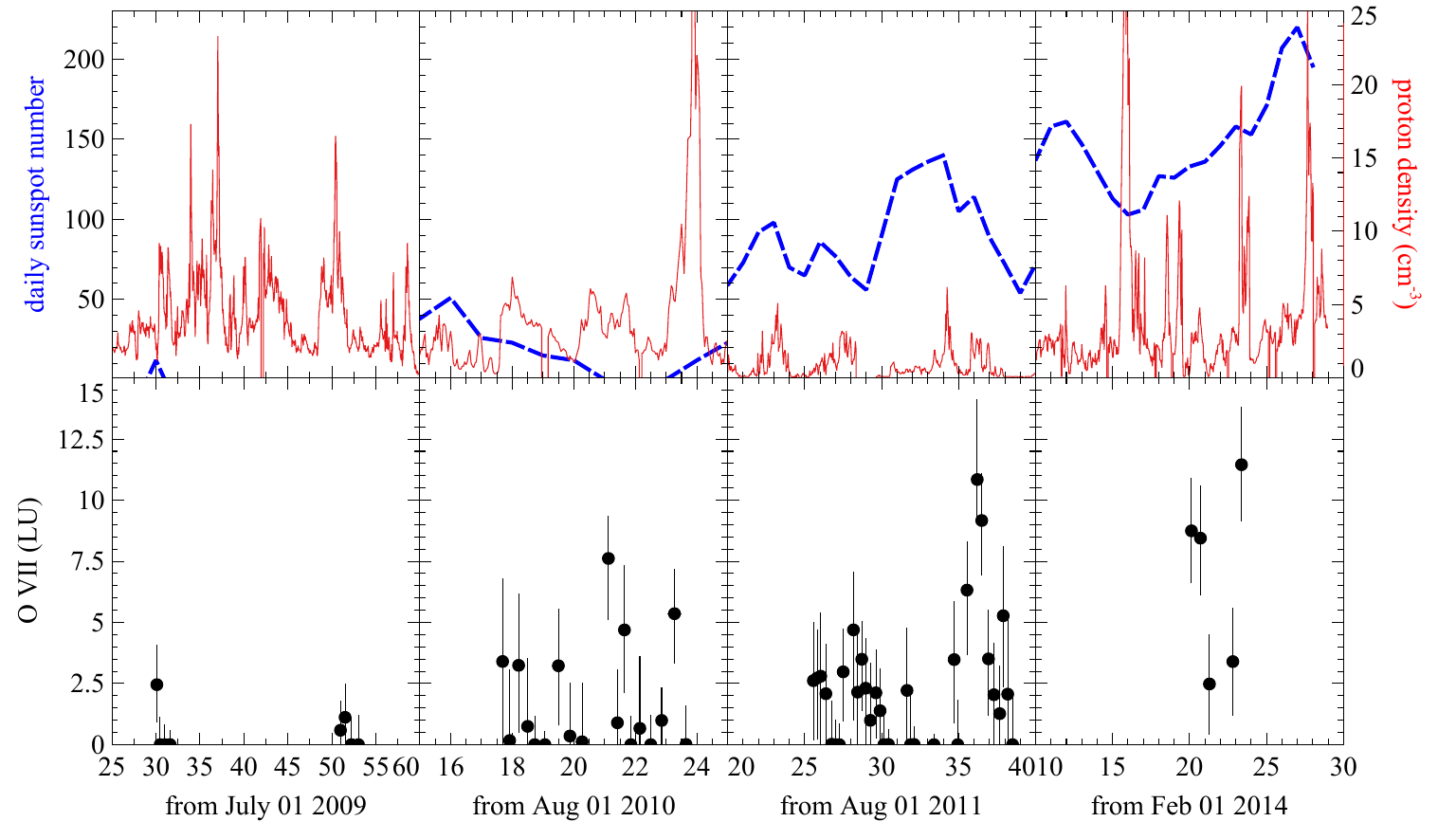}}
\caption{(upper panels) The daily sunspot number and proton density plotted against the observation date. (lower panels) The strengths of excess O\,\emissiontype {VII} line of the Perseus cluster and the NW outskirt of the AWM7 cluster.   {Alt text: Eight-panel plot arranged in two rows and four columns. The upper panels display the daily sunspot number over time for four observing periods, along with the corresponding solar wind proton density for the same periods. The bottom panels present the excess O VII line intensity measured toward the Perseus cluster and the northwestern outskirts of AWM7. Each column represents one observing window with its own time axis starting from a specific date. }  }
\label{fig:o7}
\end{figure*}

   Figure \ref{fig:o7} shows the time variation of the excess  O\,\emissiontype {VII}  He$\alpha$ line intensity, with the solar wind proton density observed by the ACE satellite and the daily sunspot number \footnote{SILSO, World Data Center—Sunspot Number and Long-term Solar Observations,
Royal Observatory of Belgium, online Sunspot Number catalog (2005--2015)
http://www.sidc.be/SILSO/}.
The overall trend of the strengths of the O\,\emissiontype {VII} line intensity depends on the average value of the sunspot number rather than on the local proton density observed with ACE.
   The Perseus cluster is close to the ecliptic plane (ecliptic latitude $\beta=22^\circ$),
   and the observations may be affected by the heliospheric SWCX.
   There is a hint of variation with a time scale of several days in the strengths of the excess O\,\emissiontype {VII}.

\chapter{The ICM parameters with the standard background model}
\label{sec:sicm}

\begin{figure}
    \centerline{\includegraphics[width=8cm]{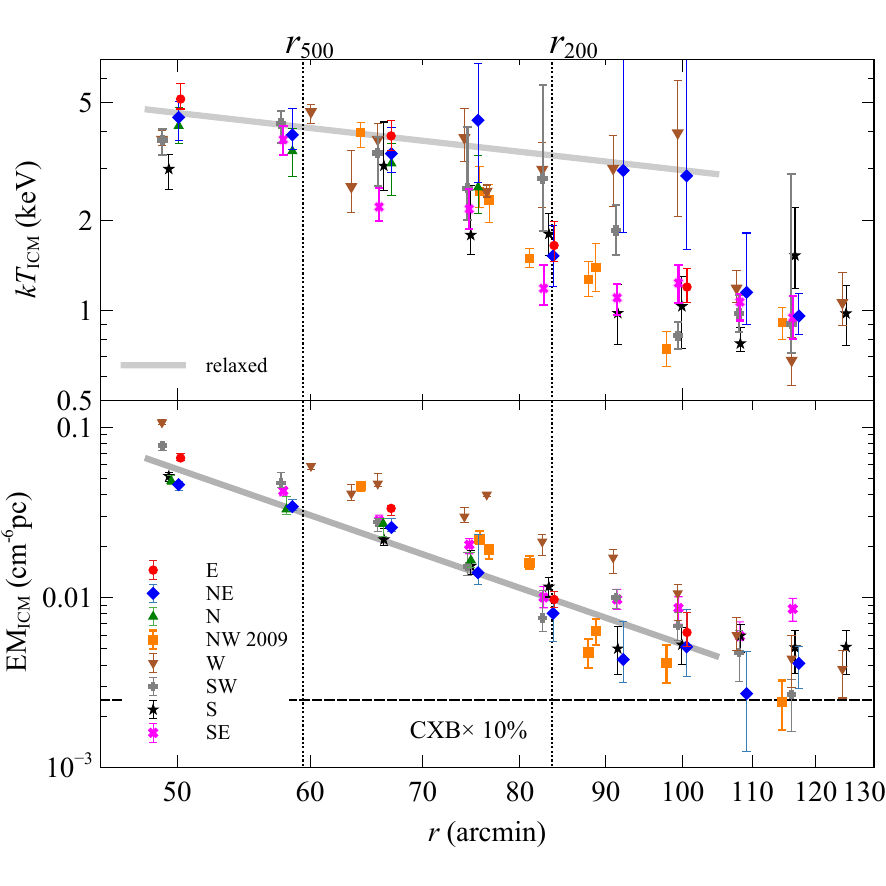}}
\caption{  The radial profiles of $kT_{\rm ICM}$ (upper panel) and EM$_{\rm ICM}$ (bottom panel) of the Perseus cluster with the standard background model fits without the HG component.
The solid gray dashed lines represent the best-fit power-law relations for the relaxed arms using the baseline model fits. The two vertical dotted lines indicate $r_{500}$ and $r_{200}$.
{Alt text: Two-panel plot showing radial profiles of the ICM temperature (top) and emission measure (bottom) in the Perseus cluster, derived from the baseline model fits. 
Different arms are represented by distinct symbols. }}
\label{fig:sicm}
\end{figure}

Figure \ref{fig:sicm} shows the radial profiles of $kT_{\rm ICM}$ and EM$_{\rm ICM}$ derived using the standard background fits (section \ref{sec:sys}). 
Compared to the baseline model fits, $kT_{\rm ICM}$ is significantly lower and becomes flat at $\sim$1 keV beyond $\sim$100$^\prime$, where
 no significant temperature differences are observed among the arms.
The EM$_{\rm ICM}$ values tend to be higher than those with the baseline model fits and also become flat beyond $\sim$100$^\prime$. 
 In contrast to the temperature profiles,  the SE arm exhibits EM$_{\rm ICM}$ values
that are higher than those on the other arms by a factor of 2--3.

\end{document}